\documentclass[11pt]{article}

\usepackage{amssymb, amsmath}
\usepackage{graphicx}
\usepackage{cite}

\newcommand{\const}{\,{\rm const}\,}

\newcommand{\cF}{\mathcal{F}}
\newcommand{\cA}{\mathcal{A}}
\def\be{\begin{equation}}
\def\ee{\end{equation}}
\def\bea{\begin{eqnarray}}
\def\eea{\end{eqnarray}}

\title{ \bf{Static near-horizon geometries in five dimensions}}

\author{Hari K. Kunduri$^a$\footnote{h.k.kunduri@damtp.cam.ac.uk } \  and James Lucietti$^b$\footnote{james.lucietti@durham.ac.uk } \\ \\
\small \sl $^a$DAMTP, University of Cambridge \\
\small \sl Centre for Mathematical Sciences, Wilbeforce Road, Cambridge, CB3 0WA, UK
\\ \\ \small \sl $^b$Centre for Particle Theory, Department of Mathematical Sciences, \\ \small \sl University of Durham, South Road, Durham, DH1 3LE, UK
 }

\date{}

\begin{document}

\maketitle

\begin{picture}(0,0)(0,0)
\put(350, 240){DCPT-09/41}
\put(350, 225){DAMTP-2009-49}
\end{picture}

\vskip1.5cm
\begin{abstract}
We consider the classification of static near-horizon geometries of
stationary extremal (not necessarily BPS) black hole solutions of
five dimensional Einstein-Maxwell theory coupled to a Chern-Simons
term with coupling $\xi$ (with $\xi=1$ corresponding to
supergravity). Assuming the black holes have two rotational
symmetries, we show that their near-horizon geometries are either
the direct product $AdS_3 \times S^2$ or a warped product of $AdS_2$
and compact 3d space. In the $AdS_2$ case we are able to classify
all possible near-horizon geometries with no magnetic fields. There
are two such solutions: the direct product $AdS_2 \times S^3$ as
well as a warped product of $AdS_2$ and an inhomogeneous $S^3$. The
latter solution turns out to be near-horizon limit of an extremal
Reissner-Nordstr\"om black hole in an external electric field.  In
the $AdS_2$ case with magnetic fields, we reduce the problem (in all
cases) to a single non-linear ODE. We show that if there are any
purely magnetic solutions of this kind they must have $S^1 \times
S^2$ horizon topology, and for $\xi^2 <1/4$ we find examples of
solutions with both electric and magnetic fields.

\end{abstract}

\newpage

\section{Introduction}
The classification of all stationary black holes in higher dimensions remains a challenging open problem. A number of general results have now been proved which constrain the space of solutions. For example, in the case of asymptotically flat black holes, it has been shown that spatial sections of the horizon must be of positive Yamabe type~\cite{GS, Galloway}; in five dimensions this implies $S^3$ or $S^1 \times S^2$ topology. It has also been shown that the event horizon must be a Killing horizon of a Killing vector field $V$~\cite{HIW} (at least in the non-extremal case\footnote{For partial results in the extremal case, see~\cite{HI}.}). Thus the classification splits into two classes, depending on whether $V$ is proportional to the stationary Killing field or not, which are referred to as the non-rotating and rotating cases respectively.

In the rotating case one can further prove the so-called rigidity theorem: i.e. there exists at least one rotational $U(1)$ isometry, such that $V=\partial_t+ \sum_{i=1}^N\Omega_i m_i$ where $m_i$ are set of commuting rotational Killing fields such that $N \leq [(D-1)/2]$ and the constants $\Omega_i$ are the angular velocities (of the horizon relative to infinity)~\cite{HIW,IM}. Most results on rotating black holes to date have assumed that $N=[(D-1)/2]$, since, as in $D=4$, in $D=5$ this renders the Einstein equations integrable. This has allowed many of the methods used in $D=4$, such as solution-generating methods, to be extended to $D=5$.

Asymptotically flat non-rotating non-extremal black holes of Einstein-Maxwell
theory can be shown to be static~\cite{SW}. Their classification thus reduces
to determining all static solutions. In the vacuum this has been
accomplished, proving the uniqueness of higher dimensional
Schwarzschild within the class of asymptotically flat, vacuum,
static black holes~\cite{GISvac}. In Einstein-Maxwell theory it has been shown
that the Reissner-Nordstr\"om black hole is the unique asymptotically
flat static black hole assuming
there are no magnetic fields (i.e. the gauge field is purely electric)~\cite{GISEM}. In four dimensions, without loss of
generality one can dualise between electric and magnetic fields to
set the magnetic field to zero. However, in higher dimensions this
is no longer the case and thus the classification of asymptotically
flat, non-extremal static black holes in higher dimensional Einstein-Maxwell
theory is incomplete.

A further complication arises in odd dimensions as one can couple
this theory to a Chern-Simons (CS) term. In fact, for a special value of
the CS coupling this theory is the bosonic sector of minimal
ungauged supergravity, which can be embedded straightforwardly in string theory. It
should be pointed out when one has a CS term, non-rotating does \emph{not}
imply static (e.g. BMPV~\cite{BMPV} and the supersymmetric black ring~\cite{susyring}), and thus the
classification problem in such theories would require even more
work.

It is also of interest to consider the classification problem for extremal black holes. In four dimensions, asymptotically flat static extremal black hole solutions of Einstein-Maxwell theory\footnote{As far as we are aware, it has not been shown that non-rotating implies static for this class of solutions.} have been classified~\cite{CT} and in the case of a connected horizon the only possibility is extremal Reissner-Nordstr\"om \footnote{The general case is given by the Majunmdar-Papapetrou black hole\cite{CT}.}. It has been argued\cite{Rog1, Rog2} this result extends to higher dimensional Einstein-Maxwell theory. In any case, as in the non-extremal case, it was assumed there were no magnetic fields, and thus the classification of such black holes is an open problem.

It is the purpose of this paper to address the systematic study of
extremal (not necessarily supersymmetric) black holes in five dimensional Einstein-Maxwell theory allowing for a CS term with coupling $\xi$. In our analysis we will
consider arbitrary $\xi$, with special focus on minimal supergravity
$\xi=1$ and pure Einstein-Maxwell ($\xi=0$). The restriction to
extremal black holes allows us to further simplify the problem by
considering the classification of their near-horizon geometries. In
fact for $\xi=1$ the classification of near-horizon geometries of
supersymmetric black holes was previously achieved in~\cite{R}. We will thus be
concerned with non-supersymmetric extremal black holes. In fact, the
classification of 5d extremal vacuum black holes with $R \times
U(1)^2$ was recently achieved~\cite{KLvac}, and these are also solutions to the
theory in question with vanishing Maxwell field. Hence, it remains
to classify non-supersymmetric near-horizon geometries with a
non-trivial Maxwell field -- these can be thought of as interpolating between the supersymmetric ones and the vacuum ones. It turns out this is a significantly more
challenging problem. We will therefore restrict attention to static
near-horizon geometries in this paper and in a future article
consider the remaining more general set of stationaty but non-static near-horizon
geometries. Note that static near-horizon geometries can arise as
near-horizon limits of both static and stationary non-static extremal black
holes~\cite{KLR2}.

The analogous problem in four dimensions has previously been solved~\cite{CT,KL4d} (and in fact the non-static axisymmetric \cite{KL4d,LP} and the
supersymmetric \cite{CRT2} cases have also been solved). The result is that the most general static near-horizon geometry (with spatially compact horizon sections) of Einstein-Maxwell theory is the simple direct product AdS$_2\times S^2$. This is, of course, the near-horizon limit of asymptotically flat extremal Reissner-Nordstr\"om. In five dimensions there exist two obvious generalisations of this: the direct products AdS$_2 \times S^3$ and AdS$_3 \times S^2$ (spatial compactness of the horizon requires one to take a quotient of the AdS$_3$ case). These are solutions for all values of the CS coupling $\xi$ and are in fact both maximally supersymmetric in minimal supergravity ($\xi=1$) just as AdS$_2\times S^2$ is maximally supersymmetric in 4d~\cite{KPTod}. Furthermore they also both arise as near-horizon limits of known asymptotically flat extremal black holes: 5d electric Reissner-Nordstr\"om (for any $\xi$) and the supersymmetric black ring ($\xi=1$) respectively. The question this paper will address is whether there are any other static near-horizon geometries (with spatially compact horizons) in such five dimensional theories. We find that this is indeed the case.

One source of complication, mentioned above, is that in five dimensions the electric and magnetic fields are not related by duality transformations as in four dimensions. Therefore a full treatment requires considering both fields. Surprisingly, even in the absence of a magnetic field the situation is more complicated than in four dimensions. We have classified such near-horizon geometries assuming the full black hole has $R \times U(1)^2$ isometry. It turns out that in addition to AdS$_2 \times S^3$ one can also have another solution which is a warped product of AdS$_2$ and an inhomogeneous $S^3$. This solution in fact corresponds to the near-horizon limit of extremal RN in a background electric field (which can be generated by a Harrison transformation). The analogous solution in 4d suffers from conical singularities.

We also find that the classification problem is substantially more challenging with magnetic fields turned on. We are not able to provide a general classification of static near-horizon geometries even assuming rotational symmetries. However we can show that in the purely magnetic case one must have $S^1 \times S^2$ horizon topology. One such solution is of course AdS$_3\times S^2$.  The other possibilities correspond to (possibly) warped products of AdS$_2$ and $S^1 \times S^2$. We find evidence for the non-existence of such solutions. The final possibility corresponds to having \emph{both} electric and magnetic fields present. For $0 \leq \xi^2 < 1/4$ we find some examples. In pure Einstein-Maxwell there is the simple direct product AdS$_2 \times S^1 \times S^2$ which is the near-horizon limit of an extremal black string (this can be constructed by oxidising the $Q=\pm P$ RN to 5d). For $0< \xi^2<1/4$ the examples we find all have $S^3$ horizon topology.  We find no examples of this kind for $\xi^2 \geq 1/4$.

This paper is organised as follows. In Section 2 we present the near-horizon equations to be solved, which arise from decomposing the Einstein-Maxwell-CS equations on spatial cross sections of the horizon. In section 2 we also define an invariant notion of the electric and magnetic fields. Section 3 contains general results on static near-horizon geometries and in Section 4 we focus on static geometries with $U(1)^2$ rotational symmetry. Section 5 provides a summary of our results. Finally, we provide an Appendix which contains the near-horizon geometry of extremal Reissner-Nordstr\"om immersed in a background electric field.

\section{Near-horizon equations}
\label{sec:NHequations}
We will consider solutions of $D=5$ Einstein-Maxwell theory coupled to a Chern-Simons term. The field equations are
\begin{eqnarray}\label{EOM1}
&& R_{\mu\nu} = T_{\mu\nu} \equiv 2\left(\cF_{\mu}^{\phantom{a}\delta}\cF_{\nu \delta} - \frac{1}{6}g_{\mu\nu}\cF^2\right)  \\ \label{EOM2}
&&d\star \mathcal{F} +\frac{2 \xi}{\sqrt{3}}\cF \wedge \cF = 0, \qquad d\cF =0
\end{eqnarray}
where $\cF$ is the Maxwell two form and we write $\cF=d\cA$.
We will be mainly interested in the cases $\xi=0$ and $\xi=1$ which correspond to pure Einstein-Maxwell and the bosonic sector of minimal ungauged supergravity respectively.

We will assume that the event horizon of a five dimensional stationary extremal black hole solution must be a Killing horizon of a Killing vector field $V$ (see~\cite{HI} for partial results). In a neighbourhood of such a Killing horizon we can always introduce Gaussian  null coordinates $(v,r,x^a)$ such that $V = \partial / \partial v$, the horizon is at $r=0$ and $x^a$ ($a=1,2,3$) are coordinates on $\mathcal{H}$, a spatial section of the horizon. We will assume that $\mathcal{H}$ is a three-dimensional compact manifold without boundary. The black hole metric and Maxwell field in these coordinates are
\bea
 ds^2 &=& r^2 F(r,x) dv^2 + 2dvdr + 2rh_a(r,x) dvdx^a + \gamma_{ab}(r,x)dx^a dx^b \\
\label{MaxGau}
\cF &=& \cF_{vr}dv\wedge dr + \cF_{ra}dr \wedge dx^{a}  + {\cF}_{va} dv \wedge
dx^a  +\frac{1}{2}\cF_{ab}dx^{a} \wedge dx^{b} \; .
\eea
The near-horizon limit \cite{R,KLR2} is obtained by taking the limit $v \to v/\epsilon, \ r \to \epsilon r$ and $\epsilon \to 0$. The resulting metric is
\begin{equation}\label{gNH}
 ds^2 = r^2F(x)dv^2 + 2dvdr + 2rh_a(x)dv dx^a + \gamma_{ab}(x)dx^a dx^b
\end{equation} where $F, \ h_a, \ \gamma_{ab}$ are a function, a one-form, and a Riemannian metric respectively, defined on $\mathcal{H}$. In general the Maxwell field~(\ref{MaxGau}) does \emph{not} admit a near-horizon limit due to the $\cF_{va}$ component. However, we can use the field equations to show that it must for solutions of the field equations. It is well known that for a  Killing horizon $\mathcal{N}$ of $\xi$ one must have $R_{\mu\nu}\xi^{\mu}\xi^{\nu}|_{\mathcal{N}}=0$. Taking $\mathcal{N}$ to be the event horizon with $\xi=V$, and using~(\ref{EOM1}) one finds:
\be
R_{\mu\nu}\xi^{\mu}\xi^{\nu}|_\mathcal{N}= 2\gamma^{ab}
\cF_{va}\cF_{vb}|_{r=0}
\ee which implies $\cF_{va}=0$ at $r=0$. It follows (assuming analycity) that
$\cF_{va}=r \hat{\cF}_{va}$ for some regular functions
$\hat{\cF}_{va}$. This guarantees that the near-horizon limit of the Maxwell field always exists, and is given by:
\be
\cF=\cF_{vr}(x)dv \wedge dr + r\hat{\cF}_{va}(x) dv \wedge dx^a +
\frac{1}{2}\cF_{ab}(x)dx^a \wedge dx^b  \; .\ee
Note that the Bianchi identity $d\cF=0$ further constrains the Maxwell field and implies it can be written as
\be\label{FNH}
\cF=\frac{\sqrt{3}}{2}d(\Delta(x) rdv)+ \hat{\cF}
\ee
where $\hat{\cF}\equiv \frac{1}{2}\cF_{ab}(x)dx^a \wedge dx^b$ is a closed two-form and $\sqrt{3} \Delta/2 \equiv -\cF_{vr}$ is a function, both defined on $\mathcal{H}$. Note that we can locally introduce a potential $\hat{A}$ on $\mathcal{H}$ such that $\hat{\cF}=d\hat{A}$.
\par We would like to determine all possible extremal black hole near-horizon geometries in the above theory. This is equivalent to finding the most general metric and Maxwell field of the form~(\ref{gNH}) and (\ref{FNH}) that satisfy~(\ref{EOM1}) and (\ref{EOM2}). A lengthy calculation reveals that the spacetime field equations are equivalent to the following set of equations on $\mathcal{H}$:
\begin{eqnarray}
 R_{ab} &=& \frac{1}{2}h_a h_b - \nabla_{(a}h_{b)}  + 2\hat{\cF}_{ac}\hat{\cF}_{bd}\gamma^{cd} + \frac{1}{2}\Delta^2\gamma_{ab} - \frac{\gamma_{ab}}{3}\hat{\cF}^2 \label{Riceq} \\
F &=& \frac{1}{2}h_ah^a - \frac{1}{2}\nabla_a h^a -\Delta^2 - \frac{\hat{\cF}^2}{3} \label{Feq} \\
\label{maxeq}
d\star_3 \hat{\cF} &=& -\star_3 i_h\hat{\cF}- \frac{\sqrt{3}}{2} \star_3 (d\Delta-\Delta h)+2 \xi \Delta \hat{\cF}.
\end{eqnarray} where $R_{ab}$, $\nabla$ and $\star_3$ are the Ricci tensor, the covariant derivative and Hodge dual of the 3d metric $\gamma_{ab}$. In particular, (\ref{Riceq}) is the $ab$ component of the Einstein equations, (\ref{Feq}) is the $vr$ component, and (\ref{maxeq}) is the Maxwell equation, all written covariantly on $\mathcal{H}$. It can be shown that the rest of the Einstein equations are satisfied as a consequence of the above set of equations. We have not been able to solve these equations in general (indeed without the assumption of symmetries this has not been even done in the pure vacuum case~\cite{KLvac}). It will be useful for later purposes to note that the contracted Bianchi identity $\nabla^a\left(R_{ab}-\frac{1}{2}R\gamma_{ab}\right)=0$ is equivalent to
\be \nabla_a F =  Fh_a +2h^b{\nabla}_{[a}h_{b]} - {\nabla}^{b}{\nabla}_{[a}h_{b]}  -\sqrt{3}\left( \hat{\cF}_{ab}+ \frac{\sqrt{3}}{2} \Delta \gamma_{ab} \right)( \nabla^b \Delta- h^b\Delta ) \label{gradF} \; .
\ee
In this paper we will focus on solving the near-horizon equations for {\it static} near-horizon geometries.

Before moving on it will be useful to introduce an invariant notion of electric and magnetic fields associated with the Maxwell field $\cF$. We define\footnote{Typically one defines the electric and magnetic fields with respect to the stationary Killing field rather than the co-rotating one. Of course for non-rotating black holes there is no difference.}  the {\it electric field} $\mathcal{E}$, which is a 1-form, by
\be
\mathcal{E}=-i_V \cF
\ee
and the {\it magnetic field} $\mathcal{B}$, which is a 2-form, by
\be
\mathcal{B}=i_V \star \cF \; .
\ee
In fact, the Maxwell field may be completely reconstructed from the fields $\mathcal{E}, \mathcal{B}$:
\be
\cF=-\frac{1}{V^2} [V \wedge \mathcal{E} +\star(V \wedge \mathcal{B}) ] \; .
\ee
Using the invariance of the Maxwell field $\mathcal{L}_V \cF=0$ and the Bianchi identity it follows that $d\mathcal{E}=0$ and therefore locally in the spacetime one can always introduce the electric potential $\Phi$ such that $\mathcal{E}=d\Phi$. For $\xi=0$ a similar argument shows that $\mathcal{B}=d\Psi$ for some 1-form potential $\Psi$ which we call the magnetic potential. For a near-horizon geometry we may compute the electric and magnetic fields in terms of the near-horizon data $(F,h_a,\gamma_{ab},\Delta, \hat{\cF})$. We find
\be
\mathcal{E} = \frac{\sqrt{3}}{2} d(r\Delta)
\ee
and therefore
\be
\Phi=\frac{\sqrt{3}}{2} r\Delta +\Phi_H
\ee
where $\Phi_H$ is a constant corresponding to the electric potential on the horizon. Also we find
\be
\mathcal{B}= dr \wedge \star_3 \hat{\cF}- r\left( \frac{\sqrt{3}}{2} \star_3(d\Delta-h\Delta) +\star_3 i_h \hat{\cF} \right) = d(r \star_3 \hat{\cF})-2\xi r \Delta \hat{\cF}
\ee
where the second equality follows from the near-horizon Maxwell equation (\ref{maxeq}). Therefore when $\xi=0$ we have $\Psi=r \star_3\hat{\cF} +d\lambda$ for some function $\lambda$. The expression for $\mathcal{E}$ and $\mathcal{B}$ allow us to deduce the following important facts for a near-horizon geometry:
\begin{itemize}
\item A near-horizon geometry has vanishing electric field (i.e. $\mathcal{E}\equiv 0$) if and and if $\Delta \equiv 0$.
\item A near-horizon geometry has vanishing magnetic field (i.e. $\mathcal{B} \equiv 0$) if and only if $\hat{\cF} \equiv 0$.
\end{itemize}

\section{Static near-horizon geometries}
A static near-horizon geometry is defined as one for which the
Killing field normal to the event horizon is hypersurface
orthogonal, i.e. $V \wedge dV =0$ everywhere. In~\cite{KLR2} it was
shown that in terms of the metric near-horizon data this is
equivalent to $dF=hF$ and $dh=0$.  One can derive an analogous
constraint for the Maxwell field~\cite{KL4d}. Defining the twist one-form
$\omega = \frac{1}{2} \star(V \wedge dV)$, one can check $d\omega =
-\star(V \wedge R(V))$ where $R(V)_{\mu} = R_{\mu\nu}V^{\nu}$.
Therefore a static near-horizon must be Ricci-static, i.e. $V \wedge
R(V)=0$. One can check that a near-horizon geometry is Ricci-static if and only if $d\Delta =  h\Delta$. Therefore any static near-horizon geometry must satisfy
$d\Delta =h \Delta$.

We will solve these staticity conditions
assuming $\mathcal{H}$ is a closed manifold -- we will see that
their solution depends on the topology of $\mathcal{H}$. Since
$dh=0$, Hodge's decomposition theorem tell us we can write
$h=\beta+d\lambda$ where $\beta$ is a globally defined harmonic
1-form (equivalently $d\beta=0=\nabla_a\beta^a$), $\lambda$ is a
globally defined function. The analysis splits into two depending on whether $\beta \equiv 0$ or not. If $H^1(\mathcal{H})=0$ then there are no harmonic 1-forms so $\beta \equiv 0$. However note that if $H^1(\mathcal{H}) \neq 0$ one could still have $\beta \equiv 0$.

\subsection{AdS$_2$ near-horizon geometries}
In this section we analyse the $\beta \equiv 0$ case (which includes the $H^1(\mathcal{H})=0$ case); we will show that the corresponding near-horizon geometry is always a (possibly warped) product of AdS$_2$ with $\mathcal{H}$.

Since $\beta \equiv 0$ we have $h=d\lambda$. The staticity conditions are easily solved to give $F=F_0\psi^{-2}$
and $\Delta=e\psi^{-2}$ where we have defined the globally defined
function $\psi= \exp(-\lambda/2)$ and $F_0,e$ are constants of intergration.
The general near-horizon equations in this case simplify and can be
written as: \bea
&& \psi R_{ab}= 2\nabla_a\nabla_b \psi + \frac{e^2\gamma_{ab}}{2\psi^3}+ \psi \left(2\hat{\cF}_{ac}\hat{\cF}_{bd}\gamma^{cd}  - \frac{\gamma_{ab}}{3}\hat{\cF}^2 \right) \label{Ricstatic}\\
&& F_0 =\frac{1}{2}\nabla^2\psi^2 -\frac{e^2}{\psi^2}-\frac{\psi^2\hat{\cF}^2}{3} \label{F0} \\
&&d ( \psi^2 \star_3 \hat{\cF})-2\xi e \hat{\cF}=0
\label{maxstatic}\; . \eea In the pure vacuum case $e=0$ and
$\hat{\cF} \equiv 0$, assuming compactness of
$\mathcal{H}$ greatly simplifies the problem and it has been shown this implies $\psi= \const$, $F_0=0$ and
$R_{ab}=0$ -- this corresponds to a near-horizon geometry $R^{1,1}$
times a Ricci flat 3d space~\cite{CRT1}. We will focus on solutions to these
equations with non-zero spacetime Maxwell field, i.e. such that not
both $e=0$ and $\hat{\cF}\equiv 0$. Notice that integrating
(\ref{F0}) over $\mathcal{H}$ then implies that $F_0<0$ and thus we can introduce a positive constant $C$ such that $F_0=-C^2$. The near-horizon geometry is a (warped) product of AdS$_2$ and
$\mathcal{H}$. This can be seen by defining a new radial variable $r \to \psi^2r$ in terms of which the full near-horizon geometry is
\be
ds^2=\psi^2( -C^2r^2 dv^2+2dvdr)+ \gamma_{ab}dx^adx^b \; .
\ee
We will now make some general observations regarding solutions to these equations.

\subsubsection{Electric case}
Firstly
note that in the case of vanishing magnetic field (i.e. $\hat{\cF}
\equiv 0$) and $\psi =\const$ everywhere (i.e. $h \equiv 0$), then
$\mathcal{H}$ is a positive curvature Einstein space and thus its
metric must be locally isometric to a round $S^3$. The near-horizon
geometry is simply the familiar solution $AdS_2 \times S^3$. One
might be tempted to conjecture that this is the only solution with
no magnetic field (as AdS$_2 \times S^2$ is in 4d). In fact we will
show there are other more non-trivial solutions of this kind by
explicitly classifying all possibilities with $U(1)^2$ symmetry.

\subsubsection{Magnetic case}
Suppose $e=0$. Then one can solve the Maxwell equation $\psi^2 \star_3 \hat{\cF}= d\chi$ for some (locally defined) function $\chi$. If $H^1(\mathcal{H})=0$ then $\chi$ is a globally defined function on $\mathcal{H}$. This allows us to show if $\psi$ is a constant then $H^1(\mathcal{H}) \neq 0$ as follows. If $\psi$ is a constant then $\hat{\cF}^2$ is constant and thus $| d\chi |^2$ is constant. But since $\chi$ is a function on a closed manifold it must have a maximum (and minimum) somewhere, i.e. $d\chi=0$ somewhere. It follows that $d\chi \equiv 0$ everywhere and thus $\hat{\cF} \equiv 0$. This shows that there are no purely magnetic solution with $\psi$ constant and $H^1(\mathcal{H})=0$ horizon topology (which includes $S^3$). Further, since when $\psi$ is constant $R=\hat{\mathcal{F}}^2>0$ (i.e. the horizon is positive Yamabe) in this case it follows that $\mathcal{H}=S^1 \times S^2$.  It would be interesting if this result could be strengthened by removing the assumption that $\psi$ is constant. In fact later by assuming $U(1)^2$ symmetry we will show that the horizon of a purely magnetic solution must be $S^1 \times S^2$ topology without assuming $\psi$ is constant (and in fact there are none with such symmetry and $\psi$ constant). We should emphasise we have not found any examples of geometries in this class and it is possible that they do not exist.

\subsubsection{Electro-magnetic case}
We have gathered some general results for when both electric and magnetic fields are present and the CS coupling $\xi \neq 0$. In particular the near-horizon Maxwell equation, together with compactness of $\mathcal{H}$, can be used to establish an interesting simplification of these equations in the case $\xi e \neq 0$. \\

\noindent {\bf Lemma 1:} Consider $\xi e \neq 0$ and assume $\hat{\cF}$ is not identically zero. Assume $\mathcal{H}$ is compact (in fact closed and orientable). Then the function $\psi$ is constant.\\

\noindent {\bf Proof:} The proof employs the Maxwell equation (\ref{maxstatic}) together with Hodge theory on $\mathcal{H}$. The Hodge theorem for $p$-forms states that the Laplacian $\Delta_H= dd^{\dagger}+ d^{\dagger} d$, where $d^{\dagger}=\star_3d\star_3$ is the adjoint of $d$ under the inner product $(\alpha,\beta)=\int_{\mathcal{H}} (\star_3 \alpha) \wedge \beta$, possesses an orthonormal set of eigen-forms $\omega_n$, so that $\Delta_H \omega_n =\lambda_n \omega_n$ (with $\lambda_n \geq 0$) and $(\omega_n, \omega_m)=\delta_{nm}$\cite{Lap}. In general the multiplicity of each eigenvalue may be greater than one, but must be finite which we denote by $\textrm{mult}(\lambda_n)$. Now,  note that (\ref{maxstatic}) can be written as $L \tilde{F}= \ell \tilde{F}$ where we have defined the operator $L \equiv d \star_3= \star_3 d^{\dagger}$, the two form $\tilde{F} \equiv \psi^2 \hat{\cF}$ and the function $\ell \equiv 2\xi e \psi^{-2}$. In fact, the operator $L$ is self-adjoint and commutes with $\Delta_H$ and thus defines a self-adjoint map on each eigenspace $V_n=\{ \omega \in \Lambda^2(\mathcal{H}) | \;\Delta_H \omega= \lambda_n \omega \}$. Since $\textrm{dim}V_n=\textrm{mult}(\lambda_n)$ is finite dimensional, $L$ can be diagonalised within it and thus $L\omega_n^i=\beta^i_n \omega_n^i$ for $i=1, \dots \textrm{mult}(\lambda_n)$ (where the eigenvalues $\beta^i_n$ are constants of course) with $(\omega_n^i, \omega^j_n)=\delta^{ij}$. Thus, expanding $\tilde{F}= \sum_{n,i} f_n^i \omega^i_n$, the equation $L\tilde{F}=\ell \tilde{F}$ is simply $f_n^i\beta^i_n= \ell f^i_n$ (no sum). Since by assumption there is at least one $f^i_n \neq 0$ (otherwise $\hat{\cF} \equiv 0$) it follows that $\ell$ is a constant. Finally, since $\ell= 2\xi e/\psi^2$ it follows that if $\xi e \neq 0$ then $\psi$ is a constant which proves the Lemma. \\

Let us persist with the case $\xi e \neq 0$. One can rewrite the Maxwell equation (\ref{maxstatic}) by defining
\be
\hat{\cA} \equiv \frac{\psi^2}{2\xi e} \star_3 \hat{\cF}
\ee
so the Maxwell equation reads $d\hat{\cA}=\hat{\cF}$ -- i.e. $\hat{\cA}$ is a globally defined potential. The near-horizon equations can be rewritten in terms of $\cA$ and for $\psi=\const$ are simply
\bea
&& R_{ab}= \frac{\Delta^2\gamma_{ab}}{2}+ 8\Delta^2\xi^2 \left( \frac{2}{3} \hat{\cA}_c \hat{\cA}^c \gamma_{ab}- \hat{\cA}_a \hat{\cA}_b\right) \label{RicA}\\
&& \label{F0A} F = -\Delta^2-\frac{8\Delta^2\xi^2 \hat{\cA}_c \hat{\cA^c}}{3} \; \\ \label{Aeq}
&& d\hat{\cA} = 2\xi \Delta \star_3 \hat{\cA} \;  \eea
where we have rewritten things in terms of the constant functions $\Delta=e \psi^{-2}$ and $F=F_0\psi^{-2}$.
From equation (\ref{F0A}) it follows that $\hat{\cA}^2$ is a constant. If
$\hat{\cA}^2=0$ then $\hat{\cF} \equiv 0$ and thus here we assume $\hat{\cA}^2>0$. Also note that $d \star_3 \hat{\cA}=0$ and thus $\Delta_H \hat{\cA}= (2\xi \Delta)^2 \hat{\cA}$. Thus $\hat{\cA}$ is an eigen 1-form of the Laplacian (or equivalently $\hat{\cF}$ is an eigen 2-form): it follows that $\hat{\cA}= \const \omega_n$ and $(2\xi \Delta )^2 =\lambda_n$ (see above proof for definitions) for some $n$.
We now
look for solutions to this system of equations (for convenience we
drop the ``hats'' on $\cA$).\\

\noindent {\bf Lemma 2:} Consider $\xi e \neq 0$ and $\psi=\const$. Assume ${\cA}^2 \neq 0$ and that $\cA^a$ is a Killing vector
field. If $\xi^2\geq 1/4$ there are no solutions to the system (\ref{RicA}), (\ref{F0A}) and (\ref{Aeq}). If $0<\xi^2<1/4$ the metric on $\mathcal{H}$ is locally isometric to a homogeneously squashed metric on $S^3$.
\\

\noindent {\bf Proof:} Introduce a
coordinate so $\cA=\partial /\partial z$ and thus the metric on
$\mathcal{H}$ is \be \gamma_{ab}dx^adx^b= \cA^2(dz+ \bar{\alpha}_a
d\bar{x}^a)^2+ \bar{g}_{ab} d\bar{x}^ad\bar{x}^b \ee where
$\bar{\alpha}$ and $\bar{g}$ are a 1-form and metric depending only
on the coordinates $\bar{x}^a$ of the 2d space formed by quotienting
by the action generated by $\cA$. The horizon equations (\ref{RicA})
reduced on this 2d space are: \bea
&&\cA^2 \bar{\nabla}_{[a} \bar{\alpha}_{b]} \bar{\nabla}^{[a} \bar{\alpha}^{b]} = \Delta^2 \left( \frac{1}{2}-\frac{8\xi^2  \cA^2}{3} \right)  \label{alpha1} \\
&&\cA^2\bar{\nabla}^b \bar{\nabla}_{[a} \bar{\alpha}_{b]} + \bar{\alpha}_a \cA^4\bar{\nabla}_{[c} \bar{\alpha}_{b]} \bar{\nabla}^{[c} \bar{\alpha}^{b]} = \Delta^2\cA^2\left( \frac{1}{2}-\frac{8\xi^2  \cA^2}{3} \right) \bar{\alpha}_a \label{alpha2} \\
&&\bar{R}_{ab} +\bar{\alpha}_{a} \cA^2\bar{\nabla}^c \bar{\nabla}_{[b} \bar{\alpha}_{c]} +\bar{\alpha}_{b} \cA^2\bar{\nabla}^c \bar{\nabla}_{[a} \bar{\alpha}_{c]}+ \bar{\alpha}_a \bar{\alpha}_b \cA^4 {\nabla}_{[c} \bar{\alpha}_{b]} \bar{\nabla}^{[c} \bar{\alpha}^{b]} \nonumber \\
&& \; = 2\cA^2\bar{g}^{cd} \bar{\nabla}_{[c} \bar{\alpha}_{a]} \bar{\nabla}_{[d} \bar{\alpha}_{b]}  +\Delta^2\cA^2\left( \frac{1}{2}-\frac{8\xi^2 \cA^2}{3} \right) \bar{\alpha}_a \bar{\alpha}_b + \Delta^2 \left( \frac{1}{2}+ \frac{16 \xi^2 \cA^2}{3} \right) \bar{g}_{ab} \label{alpha3}
\eea
where $\bar{\nabla}$ and $\bar{R}_{ab}$ are the metric connection and Ricci tensor associated to $\bar{g}_{ab}$. Also note that the Maxwell equation (\ref{Aeq}) reads $(\cA^2)^{1/2} \bar{\nabla}_{[a} \bar{\alpha}_{b]}= \xi \Delta \bar{\epsilon}_{ab}$, where $\bar{\epsilon}$ is the volume form associated with $\bar{g}$, and thus equation (\ref{alpha1}) implies $\cA^2= 3(1-4\xi^2)/(16\xi^2)$ (and equation (\ref{alpha2}) is automatic). It follows that solutions only exist for $\xi^2<1/4$. In this case the equation on the 2d space reduce to just
\be
\bar{R}_{ab}= \Delta^2 \left( \frac{1}{2}+2\xi^2 +\frac{16\xi^2\cA^2}{3}\right) \bar{g}_{ab}
\ee
and so the metric $\bar{g}_{ab}$ is locally isometric to the round metric on $S^2$. Further,  since $d\bar{\alpha}= \frac{2\xi \Delta}{(\cA)^{1/2}} \bar{\epsilon}$ we see that the metric on $\mathcal{H}$ is locally isometric to a homogeneous metric on $S^3$. \\

In view of this Lemma it is tempting to wonder whether one can show that $\cA$ must always be Killing when $\psi$ is constant. In fact for special values of the CS coupling $0<\xi^2 <1/4$ we will find examples showing this is not the case.  The results are all consistent though with the possible non-existence of near-horizon geometries with compact $\mathcal{H}$ and both electric and magnetic fields non-vanishing for $\xi^2 \geq 1/4$ (which includes minimal supergravity).\\

Finally let us comment on the $\xi=0$ theory. As in the purely magnetic case the Maxwell equation can be solved $\psi^{2}\star_3 \hat{\cF}= d\chi$ for some function $\chi$. By the same argument as in the purely magnetic case one can show that if $\psi$ is a constant then $\mathcal{H}=S^1 \times S^2$. As we will show later, by assuming $U(1)^2$ symmetry we can show this case corresponds to a direct product AdS$_2 \times S^1 \times S^2$. However, since Lemma 1 does not apply when $\xi=0$ it is unclear whether $\psi$ must be a constant and thus whether this is the most general solution with $U(1)^2$ symmetry.

\subsection{AdS$_3$ near-horizon geometries}
In this section we analyse the $\beta$ non-vanishing case so $h=\beta+d\lambda$. This case can only occur if $H^1(\mathcal{H}) \neq 0$. We should point out that we have not shown in general that such near-horizon geometry must contain an AdS$_3$ factor -- later assuming a $U(1)^2$ symmetry we do show this is the case though.

Using compactness, the staticity conditions, when $\beta$ is non-vanishing, may be completely solved. The crucial result is the following:\\

\noindent {\bf Lemma 3:} Let $\Phi$ be a function on $\mathcal{H}$
which satisfies $d\Phi=h\Phi$ and $H^1(\mathcal{H}) \neq 0$. Suppose $\beta \neq 0$ at some point $p$ in $\mathcal{H}$. Then
$\Phi \equiv 0$ everywhere on $\mathcal{H}$.\\

\noindent {\bf Proof:} Writing $h=\beta+d\lambda$ it is easy to show
that $d\Phi=h\Phi$ is equivalent to $d\tilde{\Phi}=\beta
\tilde{\Phi}$ where $\tilde{\Phi}= \exp(-\lambda) \Phi$. Since
$\nabla_a\beta^a=0$ it follows that $\nabla^2
\tilde{\Phi}=\beta_a\beta^a \tilde{\Phi}$. Since $\mathcal{H}$ is a closed
manifold this implies $\int_{\mathcal{H}}\left( |\nabla \tilde{\Phi}|^2 + \beta^2 \tilde{\Phi}^2 \right)=0$ and thus $\tilde{\Phi}$ is a constant function an $\beta^2 \tilde{\Phi}^2 \equiv 0$. Evaluating $\beta^2 \tilde{\Phi}^2$ at $p$ we deduce that $\tilde{\Phi} \equiv 0$ on $\mathcal{H}$.  \\

This result tells us that as long as $\beta$ does not vanish everywhere then the
solution to the staticity conditions is $F \equiv 0$ and $\Delta
\equiv 0$. The latter condition is equivalent to vanishing electric
field. In this case the near-horizon equations
simplify to (again defining $\psi=\exp(-\lambda /2)$):
\bea
&& \psi R_{ab}= 2\nabla_a\nabla_b \psi -2\beta_{(a} \nabla_{b)}\psi+ \psi\left ( \frac{1}{2}\beta_a\beta_b -\nabla_{(a} \beta_{b)}\right)+ \psi \left(2\hat{\cF}_{ac}\hat{\cF}_{bd}\gamma^{cd}  - \frac{\gamma_{ab}}{3}\hat{\cF}^2 \right) \label{Riceq2}\\
&& \nabla^2\psi^2 - 2 \left(d\psi^2 \cdot \beta\right) +\psi^2\left(\beta^2 -\frac{2\hat{\cF}^2}{3}\right) =0 \label{Feq2}\\
&&d ( \psi^2 \star_3 \hat{\cF})=-\psi^2 \star_3 i_\beta \hat{\cF}
\label{maxstatic2}\; . \eea We wish to solve these equations. In the pure vacuum case $\hat{\cF} \equiv 0$, integrating equation (\ref{Feq2}) over $\mathcal{H}$ (and using $\nabla_a \beta^a=0$) implies $\beta \equiv 0$ everywhere contradicting our starting assumption. Therefore there are no vacuum near-horizon geometries in this class -- this is consistent with the results of~\cite{CRT1}. We thus focus on solutions with $\hat{\cF} \neq 0$.

Using (\ref{Riceq2}) one can show that
\begin{eqnarray} R_{ab}\beta^a\beta^b &=& 2\psi^{-1}\beta^a\nabla_a( \beta \cdot d\psi)-2\psi^{-1}\beta^2 (\beta \cdot d\psi) \nonumber \\ &&-\psi^{-1}\nabla^b \beta^2 \, \nabla_b \psi + \frac{1}{2}\beta^2 \left( \beta^2-
\frac{2}{3} \hat{\cF}^2 \right) +2(i_\beta \hat{\cF})^2-
\frac{1}{2}\beta^a\nabla_a \beta^2 \end{eqnarray} which allows one to establish:
\bea
\label{lapbeta} \nabla^2 \beta^2+ \left(\beta^a+ 2 \psi^{-1}\nabla^a\psi \right)\nabla_a \beta^2 \nonumber &=&
2\nabla_a\beta_b \nabla^a\beta^b +\beta^2 \left( \beta^2- \frac{2}{3}
\hat{\cF}^2 \right) +4(i_\beta \hat{\cF})^2 \\ &&+4\psi^{-1}( \beta^a\nabla_a( \beta \cdot d\psi)-\beta^2 (\beta \cdot d\psi) ) \; .\eea This allows us
to show that if $\psi$ (i.e. $\lambda$) is constant then $\beta$ is a Killing field
(in fact covariantly constant). This can be seen as follows.
Equation (\ref{Feq2}) implies $\beta^2=\frac{2}{3} \hat{\cF}^2$.
Then (\ref{lapbeta}) tells us that $\nabla^2 \beta^2+\beta^a
\nabla_a \beta^2 \geq 0$ and hence by the maximum principle
$\beta^2$ is constant. Then (\ref{lapbeta}) implies $\nabla_a
\beta_b \equiv 0$ and $i_{\beta} \hat{\cF}\equiv 0$. This case can
then be completely solved. Noting that $d\beta=0$, so $\beta$ is
hypersurface orthogonal, we can introduce coordinates
$(z,\bar{x}^a)$ on $\mathcal{H}$ so $\beta=\partial /\partial z$ and
\be \gamma_{ab}dx^adx^b = \beta^2 dz^2+ \bar{g}_{ab}d\bar{x}^a
d\bar{x}^b,  \qquad \hat{\cF} = \frac{1}{2}\bar{F}_{ab} d\bar{x}^a \wedge
d\bar{x}^b \; .\ee We may now reduce the equations onto the 2d space
orthogonal to $\beta$. Note that we must have $\bar{F}=f
\bar{\epsilon}_2$ for some function $f$, where $\bar{\epsilon}_2$ is the volume form of $\bar{g}$. But $\beta^2=\frac{2}{3}
\bar{F}^2$ implies $f^2= 3\beta^2/4$ which is a constant. The Ricci
equation reduces to $\bar{R}_{ab}= \beta^2 \bar{g}_{ab}$ and thus
$\bar{g}_{ab}$ is simply the round metric on $S^2$. The full
near-horizon geometry in this case is \bea ds^2 = 2dvdr +2r \beta^2
dzdv + \beta^2 dz^2 + \bar{g}_{ab}d\bar{x}^a d\bar{x}^b, \qquad \hat{\cF}=
\pm \frac{\sqrt{3 \beta^2}}{2}  \bar{\epsilon}_2 \eea which is
locally isometric to the direct product $AdS_3 \times S^2$.

One might suspect that this is in fact the only solution in this class with a compact horizon although we have not been able to prove this. In fact, later we will show that this is the only regular solution with $\beta \neq 0$, compact horizon sections and a $U(1)^2$ symmetry (i.e. $\psi$ must be constant under these assumptions). It would be interesting to prove this without the assumption of $U(1)^2$ symmetry.

\section{Static near-horizon geometries with rotational symmetries}
We will now restrict consideration to near-horizon geometries of stationary
extremal black hole solutions which admit
two commuting rotational Killing vector fields. That is,
black hole solutions with an isometry group $R \times U(1)^2$ which also leaves the Maxwell field $\cF$ invariant. Denote the generators of the $U(1)^2$ isometry by $m_i$ for $i=1,2$, and
introduce coordinates adapted to these so that $m_i=
\partial /
\partial \phi^i$ with $\phi^i \sim \phi^i +2\pi $. Recall that a standard argument (e.g.~\cite{KLRnoring}) shows that $\cF(m_1,m_2)$ is constant, and furthermore this constant must be zero if at least one combination of the $m_i$ must vanish somewhere (as will be the case for the horizon topologies of interest).
The near-horizon limit of such black hole solutions inherits the $U(1)^2$ symmetry. It follows that the near-horizon data $(F,\gamma_{ab},h_a, \Delta, \hat{\cF})$ are all invariant under the $m_i$. Therefore the $m_i$ are also Killing fields of the horizon metric $\gamma_{ab}$. It turns out that the only allowed topologies of closed-3 manifolds admitting an effective $U(1)^2$ symmetry are $T^3,S^3$, $L(p,q)$ (Lens spaces) and $S^1 \times S^2$ \cite{Gowdy, Chr}. From now on we will assume non toroidal horizon topology.

This allows us to introduce canonical coordinates on the horizon as follows. Define the 1-form $\Sigma=i_{m_1}i_{m_2} \epsilon_3$, where $\epsilon_3$ is the volume form associated to the 3-metric $\gamma_{ab}$. $\Sigma$ is closed as a consequence of $m_i$ being Killing fields. It follows that for the topologies on interest, there exists a globally defined function $\sigma$ such that $\Sigma=d\sigma$~\cite{KLvac}. From these definitions one can show that $(d\sigma)^2=\det [ m_i \cdot m_j ]$ and $m_i \cdot d\sigma=0$. Therefore we learn that $d\sigma=0$ if and only if one (or a combination) of the $m_i$ vanish (i.e. $\det [m_i \cdot m_j]=0$). For the topologies we are considering this only occurs at two points on $\mathcal{H}$ which thus correspond to the minimum $\sigma_1$ and maximum $\sigma_2$ of $\sigma$. Thus $\sigma_1 \leq \sigma \leq \sigma_2$ with $d\sigma=0$ only at $\sigma=\sigma_1,\sigma_2$. Therefore we can use $\sigma$ as a coordinate for $\sigma_1<\sigma<\sigma_2$.  Collecting these results we can write the horizon metric $\gamma_{ab}$ and horizon Maxwell field $\hat{\cF}_{ab}$ in the coordinates $(\sigma,x^i)$ for $\sigma_1<\sigma<\sigma_2$
\bea \label{hor}
\gamma_{ab} dx^a dx^b  &=& \frac{d\sigma^2}{\gamma(\sigma)}+\gamma_{ij}(\sigma) dx^i dx^j, \qquad \hat{\cF} = \frac{\sqrt{3}}{2} B_i(\sigma) d\sigma \wedge dx^i
\eea
where $\gamma \equiv \det \gamma_{ij}$ and for convenience we have defined the $x^i$ as some constant linear combination of the $\phi^i$ (so $\partial /\partial x^i$ need not have closed orbits).  To analyse regularity of the metric at $\sigma=\sigma_1,\sigma_2$ will require a separate analysis. We will also use the non-negative function $Q \equiv \psi^2 \gamma$ which vanishes only when $\sigma=\sigma_1,\sigma_2$. In what follows we denote $\sigma$-derivatives by ``dots''. It remains to write the other near-horizon data $(F,\Delta, h_a)$ in these coordinates, which depends on whether $\beta \equiv 0$ or not. We thus now split the analysis into these two cases.

\subsection{AdS$_3$ near-horizon geometries}
In this section we analyse the $\beta$ non-vanishing case. Recall this occurs only if $H^1(\mathcal{H}) \neq 0$ and that we
have already shown in general that $\Delta = F = 0$ everywhere. We
have $h = \beta + d\lambda$ with $\beta$ a globally defined harmonic
one-form on $\mathcal{H}$. Since $L_{m_i}h = 0$, it follows from
uniqueness of the Hodge decomposition and the facts that
$L_{m_i}\beta$ is co-closed and $L_{m_i}d\lambda$ is exact, that
$L_{m_i}\beta=L_{m_i}d\lambda = 0$. Hence $m_i \cdot d\lambda =c_i$
for some constants $c_i$. However, since $\lambda$ must be a
periodic function of the $\phi^i$ introduced above, we must have
$c_i =0$ and hence $m_i \cdot d\lambda=0$. We will find it more
convenient to use the invariant function $\Gamma = \exp(-\lambda)> 0$ (note $\Gamma=\psi^2$).

Now, invariance of $\beta$ under the $m_i$ implies we may write
$\beta = \beta_{\sigma}(\sigma)d\sigma + \beta_i(\sigma) dx^i$
wherever the coordinates $(\sigma,x^i)$ are defined. We require
$\nabla_{a}\beta^{a} = \frac{d}{d\sigma}\beta^{\sigma} =0$ and hence
$\beta^{\sigma} = c$ for some constant $c$. Now, consider the scalar
invariant $i_\beta \Sigma = cJ$ where $J\neq 0$ is the Jacobian of
the transformation $\phi^i \to x^i$. If one of the $m_i$ vanishes
somewhere, as is the case for the topologies of interest, this shows
that $c=0$. Hence $\beta_{\sigma} = 0$. We thus have
\begin{equation}
h = \frac{k_i(\sigma)}{\Gamma(\sigma)}dx^i - \frac{\dot\Gamma(\sigma)}{\Gamma(\sigma)}d\sigma
\end{equation} where we have defined $k_i \equiv \Gamma \beta_i$. Note that $k_i$ cannot both vanish identically as otherwise $\beta \equiv 0$. Closure of $\beta$ implies $\Gamma^{-1}k_i$ are both constant.

We now turn to solving the field
equations~(\ref{Riceq2})-(\ref{maxstatic2}). Let us first consider~(\ref{maxstatic2}). This is
equivalent to the following two conditions:
\begin{equation}\label{maxstatic2b}
B_i k^i = 0 \qquad d[\Gamma \gamma^{1/2}\star_2 B] = 0
\end{equation} where $k^i = \gamma^{ij}k_j$, $B = B_idx^i$, and $\star_2$ is the Hodge star with respect to $\gamma_{ij}$, the metric induced on surfaces of constant $\sigma$. Next observe for a metric of the form~(\ref{hor}), $R_{\sigma i}$ vanishes identically. Hence from~(\ref{Riceq2}) we see $R_{\sigma i} = -\frac{1}{2}\Gamma^{-1}\gamma_{ij}\dot{k}^i = 0$ from which it follows $k^i$ are constants. It follows that $k^ik_i =C^2\Gamma $ where $C$ is a positive constant ($C$ cannot vanish as $k_i$ cannot vanish everywhere). Equation~(\ref{Feq2}) implies
\begin{equation}\label{Feq2b}
\frac{d}{d\sigma}\left(\frac{Q\dot\Gamma}{\Gamma}\right) + C^2 - QB^iB_i =0 \; ,
\end{equation}
where $B^i \equiv \gamma^{ij}B_j$.
The $ij$ components of~(\ref{Riceq2}) can be written
as
\begin{eqnarray}
 \frac{d}{d\sigma}\left(Q\gamma^{ik}\dot{\gamma}_{kj}\right) + \frac{k^ik_j}{\Gamma} + 3QB^{i}B_{j} - QB^kB_k \delta^{i}_{\phantom{i}j} = 0 \label{Ricijbeta}
\label{Ricssbeta}
\end{eqnarray}
where we have raised the $i$ index (with $\gamma^{ik}$). Taking the
trace of~(\ref{Ricijbeta})  gives
\be
\label{detads3}
\frac{d}{d\sigma} (\Gamma \dot{\gamma} ) +C^2+QB^iB_i = 0
\ee
which can be rewritten as
\begin{equation}\label{trRicij}
\ddot{Q} - \frac{d}{d\sigma}\left(\frac{Q\dot\Gamma}{\Gamma}\right) + C^2 + QB^iB_i = 0.
\end{equation} Combining~(\ref{Feq2b}) with~(\ref{trRicij}) gives
\begin{equation}\label{diffQ}
\ddot Q + 2C^2 =0
\end{equation} and hence $Q = -C^2\sigma^2 + c_1\sigma + c_2$.

Since $k^i$ are constants which cannot both vanish, we are free to
choose the $x^i$ so as to set $k^i = \delta^i_{\phantom{i}1}$. It
follows that $k^ik_i = \gamma_{11} = C^2\Gamma$. Furthermore, the
fact that $\Gamma^{-1}k_i$ are constants implies that
$\gamma_{12}/\gamma_{11}$ is constant. Hence by an appropriate shift
$x^1 \to x^1 + \textrm{const} x^2$ (which preserves $k= \partial
/\partial x^1$) we can always arrange $\gamma_{12}=0$. With these
choices the horizon metric is \be \gamma_{ab}dx^adx^b=
\frac{\Gamma}{Q} d\sigma^2+ C^2 \Gamma (dx^1)^2+ \frac{Q}{C^2\Gamma}
(dx^2)^2 \; . \ee Further, the conditions~(\ref{maxstatic2b}) imply
$B_1 = 0$ and hence using $\star_2 B = \Gamma^{-1}\gamma^{-1/2} p_i
dx^i$ for some constants $p_i$, we must have $p_2 = 0$ and
\begin{equation}
B = \frac{p}{\Gamma \gamma_{11}}dx^2
\end{equation} where $p = p_1$.

It now remains to solve for the function $\Gamma$. It turns out that
the remaining components of~(\ref{Ricijbeta}) do not yield any
further constraints.
Using the fact $\gamma_{11}=C^2\Gamma$, together with~(\ref{Feq2b})
we find that the $\sigma\sigma$ component of equation (\ref{Riceq2})
simplifies to $\ddot\Gamma=0$. Therefore we find \be \Gamma=a_0+a_1
\sigma \ee for constants $a_i$.

We have thus completely solved for the local form of the
near-horizon geometry in this case which reads (redefining the
radial coordinate $r \to \Gamma r$) \bea
&&ds^2= \Gamma[ -C^2 r^2 dv^2 +2dvdr+ C^2(dx^1+rdv)^2]+ \frac{\Gamma}{Q} d\sigma^2 + \frac{Q}{C^2 \Gamma^2} (dx^2)^2 \nonumber \\
&&\hat{\cF}= \frac{\sqrt{3}}{2}  \frac{p \; d\sigma \wedge dx^2}{ C^2
\Gamma^2} \eea which is indeed a warped product of (locally) AdS$_3$
with a 2d space $M_2$ with coordinates $(\sigma, x^2)$ and the
Maxwell field is only supported on $M_2$. We we will now show that
imposing compactness of $M_2$ implies $a_1=0$ and therefore the only
regular solution with compact $\mathcal{H}$ in this class is the
direct product of (locally) AdS$_3$ and a round $S^2$. Clearly,
$M_2$ has $S^2$ topology, with poles at the zeroes of $Q$, i.e.
$\sigma = \sigma_1,\sigma_2$.  The absence of conical singularities
at these points requires
\begin{equation}
\frac{\dot{Q}(\sigma_1)^2}{\Gamma(\sigma_1)^3} = \frac{\dot{Q}(\sigma_2)^2}{\Gamma(\sigma_2)^3}
\end{equation} and noting $\dot{Q}(\sigma_1) = -\dot{Q}(\sigma_2) = C^2(\sigma_2 - \sigma_1)$, we must have $\Gamma(\sigma_1) = \Gamma(\sigma_2)$. This in turn is equivalent to $a_1 = 0$ because the roots are distinct. Hence $\Gamma$ is constant. We now translate $\sigma$ to set $\sigma_2 = -\sigma_1$ and define new coordinates $(\theta, \phi)$:
\begin{equation}
\cos\theta = \frac{\sigma}{\sigma_1} \qquad \phi = \frac{C
\sigma_1}{\Gamma^{3/2}} x^1.
\end{equation}
It is then easy to see that $M_2$ is $S^2$ equipped with its round
metric and the near-horizon metric is a product of (locally) $AdS_3
\times S^2$:
\begin{equation}
ds^2 = \Gamma[-C^2 r^2 dv^2 + 2dvdr + C^2(dx^1 + rdv)^2] + \frac{\Gamma}{C^2}[d\theta^2 + \sin^2\theta d\phi^2].
\end{equation}

\subsection{AdS$_2$ near-horizon geometries}
Now consider the $\beta \equiv 0$ case so, as showed earlier, the near-horizon geometries in this section are all warped products of $AdS_2$ and $\mathcal{H}$. Recall that in this case $F=-C^2\Gamma^{-1}$, where $\Gamma=\psi^2=\exp(-\lambda)$ is a globally defined function. It follows that $\Gamma$ must be invariant under the $m_i$. Thus the remaining near-horizon data in coordinates $(\sigma,x^i)$ is
\bea \label{NHdataads2}
h=-\frac{\dot{\Gamma}(\sigma) d\sigma }{\Gamma(\sigma)}, \qquad F=-\frac{C^2}{\Gamma(\sigma)}, \qquad  \Delta = \frac{e}{\Gamma(\sigma)} \; .
\eea
We are now ready to write down the near-horizon equations for such near-horizon geometries. Consider the near-horizon equation (\ref{F0}). This
becomes \be \label{A0} \frac{d}{d\sigma}\left( \frac{Q \dot{\Gamma}}{\Gamma} \right)= -2C^2+ \frac{2e^2}{\Gamma}+Q B_iB_j\gamma^{ij} \ee where the Laplacian of
a function $f(\sigma)$ is \be \nabla^2 f= \frac{d}{d\sigma}\left( \frac{Q \dot{f}}{\Gamma} \right)\; .\ee
The $ij$ and $\sigma\sigma$ components of (\ref{Ricstatic}) are
\bea
&& \frac{d}{d\sigma} (Q \gamma^{ik} \dot{\gamma_{kj}}) + 3QB^iB_j + \delta^i_j \left( \frac{e^2}{\Gamma} - Q B^kB_k \right)=0\label{Ricij} \\
&&~-\frac{\ddot{Q}}{2Q} + \frac{\dot{Q}^2}{4Q^2} + \frac{\dot{\Gamma}^2}{4\Gamma^2} - \frac{\ddot\Gamma}{2\Gamma} - \frac{1}{4}\gamma^{lj}\dot\gamma_{jk}\gamma^{km}\dot\gamma_{ml} - B^iB_i -\frac{e^2}{2Q\Gamma}= 0
 \label{Ricrhorho}
\eea
respectively, where we have raised the $i$ index using $\gamma^{ij}$ and defined $B^i=\gamma^{ij}B_j$. Taking the trace of (\ref{Ricij}) gives
\be
\label{det}
\frac{d}{d\sigma} (\Gamma \dot{\gamma} ) +QB^iB_i+ \frac{2e^2}{\Gamma} = 0 \; .
\ee
In terms of $Q$ equation (\ref{det}) can be also written as
\be
\ddot{Q}- \frac{d}{d\sigma}\left( \frac{Q \dot{\Gamma}}{\Gamma} \right) + \frac{2e^2}{\Gamma} +QB_iB^i =0\label{eq2} \; .
\ee
Combining equations (\ref{A0}) and (\ref{eq2}) implies
\be
\label{Qeq}
\ddot{Q}+2C^2=0
\ee
which may be immediately integrated
\be
Q=-C^2 \sigma^2+ c_1 \sigma+ c_2
\ee
where $c_i$ are integration constants.

In order to analyse the near-horizon equations it is convenient to
single out the a particular combination of the Killing vector
fields, which without any loss of generality we take to be
$\partial/\partial x^1$. Defining $\omega= \gamma_{12}/\gamma_{11}$
the horizon metric takes the form
\be
\label{canform}
\gamma_{ab}dx^a dx^b =
\frac{\Gamma(\sigma)}{Q(\sigma)} d\sigma^2 +
\gamma_{11}(\sigma)(dx^1+\omega(\sigma) dx^2)^2 +
\frac{Q(\sigma)}{\Gamma(\sigma) \gamma_{11}(\sigma) }(dx^2)^2 \; .
\ee The $21$ component of (\ref{Ricij}) gives: \be \label{12eq}
\frac{d}{d\sigma}( \Gamma \gamma_{11}^2 \dot{\omega}) + 3\Gamma
\gamma_{11}B_1(B_2 -\omega B_1)=0 \; .
\ee
Noting that $d(Q
\gamma^{1k}\dot{\gamma}_{k1})/d\sigma =
d(Q\dot{\gamma}_{11}/\gamma_{11})/d\sigma- \Gamma
\gamma_{11}^2\dot{\omega}^2-\omega d(\Gamma \gamma_{11}^2
\dot{\omega})/d\sigma$, the $11$ component of (\ref{Ricij}) can be
simplified using (\ref{12eq}) to give \be \label{11eq} \gamma_{11}^2
\dot{\omega}^2=  \frac{1}{\Gamma} \frac{d}{d\sigma} \left(
\frac{Q\dot{\gamma}_{11}}{\gamma_{11}} \right) +
\frac{e^2}{\Gamma^2} -\frac{Q B_iB^i}{\Gamma}+
\frac{3QB_1^2}{\Gamma\gamma_{11}} \ee and using the identity
\begin{equation}
\gamma^{lj}\dot\gamma_{jk}\gamma^{km}\dot\gamma_{ml}  \equiv
\frac{(\dot\gamma_{11})^2}{(\gamma_{11})^2} +
\left(\frac{\dot\gamma_{11}}{\gamma_{11}} -
\frac{\Gamma}{Q}\frac{d}{d\sigma}\left(\frac{Q}{\Gamma}\right)\right)^2
+ \frac{2\Gamma\gamma_{11}^2\dot\omega^2}{Q}
\end{equation} where $\omega \equiv \gamma_{12}/\gamma_{11}$,
equation (\ref{Ricrhorho}) can be written as \be \gamma_{11}^2
\dot{\omega}^2= \frac{2C^2}{\Gamma} + \dot{\gamma}_{11}
\frac{d}{d\sigma} \left( \frac{Q}{\gamma_{11} \Gamma} \right) +
\frac{\dot{Q}\dot{\Gamma}}{\Gamma^2}- \frac{Q
\ddot{\Gamma}}{\Gamma^2} -\frac{e^2}{\Gamma^2}
-\frac{2QB_iB^i}{\Gamma} \; . \ee Combining these equations to as to
eliminate $\dot{\omega}^2$, and using (\ref{A0}), gives \be
\label{gamma11ode} \frac{d}{d\sigma} (\Gamma \dot{\gamma}_{11} ) +
\left( 2\ddot{\Gamma}- \frac{\dot{\Gamma}^2}{\Gamma}
\right)\gamma_{11} + 3 \Gamma B_1^2 =0 \; . \ee

The Maxwell equation (\ref{maxstatic})
reduces to solving just : \be \label{Beq} \frac{d}{d\sigma} (\Gamma \sqrt{\gamma}\star_2
B)+2\xi e B=0 \; .\ee
If $\xi=0$ one
can solve this in general to get $\star_2 B= \Gamma^{-1}\gamma^{-1/2}p_idx^i$ for constants $p_i$. For $\xi \neq 0$ it is more difficult to solve this equation
in general. However there are some solutions which
can be immediately deduced by inspection. One class is clearly given
by $B_i \equiv 0$ for any $e$. Another is given by $\star_2 B=
\Gamma^{-1}\gamma^{-1/2}p_idx^i$ (for constants $p_i$) and $e=0$.
Before considering these cases we note the following\\

\noindent {\bf Lemma 4:} Consider $\xi \neq 0$. Suppose one (or a
combination) of the rotational Killing fields is hypersurface
orthogonal. Then either
$\Delta \equiv 0$ or $B \equiv 0$. \\

\noindent {\bf Proof:} Without loss of generality we may choose the
hypersurface orthogonal Killing vector to be $\partial /\partial
x^1$. We then have $\dot{\omega}=0$ and can thus arrange $\omega=0$
by suitable shifts $x^1 \to x^1 +\const x^2$. We now see that
(\ref{12eq}) reduces to $B_1 B_2=0$. Therefore either $B_1 \equiv 0$
or $B_2 \equiv 0$. Since $\partial /\partial x^2$ is also
hypersurface orthogonal there is no loss of generality in choosing
$B_2 \equiv 0$. Now, the $x^1$ component of (\ref{Beq}) tells us
$eB_1=0$. It follows that either $\Delta=0$ or $B=0$ as claimed.\\

It is tempting to conjecture that static AdS$_2$ near horizon
geometries in theories with $\xi \neq 0$ cannot have both electric
and magnetic fields which are non-vanishing. In fact as will see
that this result is not true at least for $0<\xi^2<1/4$.

Before moving on, let us recall global constraints which arise from regularity and compactness of $\mathcal{H}$. Since the function $\sigma$ is a globally defined and non-constant, it must have distinct maxima and minima on $\mathcal{H}$ where $d\sigma=0$. From the definition of $\sigma$ such points corresponds to fixed points of the rotational Killing fields. Thus we have $\sigma_1<\sigma_2$ with $\sigma_1 \leq \sigma \leq \sigma_2$. Since $(d\sigma)^2= Q/\Gamma$ we learn that $Q>0$ for $\sigma_1 < \sigma < \sigma_2$ and $Q=0$ at $\sigma=\sigma_1,\sigma_2$. Since we have seen that $Q$ is in general a quadratic we can write $Q=C^2(\sigma_2-\sigma)(\sigma-\sigma_1)$. Note that this implies $d\sigma$ can only vanish at the two points $\sigma=\sigma_1,\sigma_2$. This is consistent with all the topologies $S^3$, $L(p,q)$ and $S^1\times S^2$. It is worth noting that the vector field $S= Q \partial /\partial \sigma$ is globally defined and vanishes at $\sigma=\sigma_1,\sigma_2$ and will prove useful for analysing regularity at these special points where $\sigma$ is not a valid coordinate.

\subsubsection{No magnetic field}\label{inhom}
This corresponds to $B_i \equiv 0$. In this case (\ref{A0}) further simplifies to
\be
\label{k0B0eq}
\Gamma\frac{d}{d\sigma}\left( \frac{Q \dot{\Gamma}}{\Gamma} \right) +2C^2\Gamma-2e^2 =0
\ee
which is very similar to the kind of equation encountered when solving for 4d near-horizon geometries, see~\cite{KLvac,KL4d}, and it can be solved by using the same technique. This consists of differentiating $\Gamma$ times the above equation and then subtracting from this $\dot{\Gamma}$ times the above equation. The result is all the terms involving $e$ cancel and using (\ref{Qeq}) one finds
\be
Q \frac{d^3 \Gamma}{d\sigma^3} +\left( \dot{Q}- \frac{Q \dot{\Gamma}}{\Gamma} \right)\left( 2\ddot{\Gamma}-\frac{\dot{\Gamma}^2}{\Gamma} \right) =0
\ee
which can be integrated to
\be
Q^2 \left( \frac{2 \ddot{\Gamma}}{\Gamma} - \frac{\dot{\Gamma}^2}{\Gamma^2} \right) = \const \; .
\ee
In fact compactness of the horizon implies this constant must vanish. This can be seen by evaluating the above at the roots of $Q$ and write the above in terms of invariants constructed from the vector field $S$~\cite{KLvac}. Therefore we have
\be
2 \ddot{\Gamma} - \frac{\dot{\Gamma}^2}{\Gamma} =0 \; .
\ee
There are two classes of solution to this ODE: (i) $\Gamma$ is a constant, and (ii) $\Gamma= \beta \sigma^2/4$ where $\beta$ is some positive integration constant.\\

\noindent {\bf Round $S^3$ horizon} \\

Now consider case (i) with $\Gamma$
constant. In this case (\ref{k0B0eq}) tells us that $C^2\Gamma=e^2$.
Equation (\ref{gamma11ode}) implies \be \gamma_{11}=b_1+b_2\sigma \ee for constants $b_i$. Substituting into
(\ref{11eq}) gives \be \dot{\omega}^2= \frac{1}{\gamma_{11}^4\Gamma}
( b_2 c_1 b_1-b_2^2 c_2 +C^2 b_1^2) \ee and therefore
$K\equiv b_2 c_1 b_1-b_2^2 c_2 +C^2 b_1^2 \geq 0$. Observe
that the equation (\ref{12eq}) is now automatically satisfied. It
remains to integrate from $\omega$; it turns out that this splits the analysis into two cases depending on whether $b_2=0$ or not.

Consider $b_2 \neq 0$ which gives \be \omega =
\frac{\sqrt{K}}{\Gamma^{1/2}b_2(b_1+b_2 \sigma)} \ee where we
have set the integration constant to zero and fixed the sign, as this can always be
done by shifting $x^1 \to \pm(x^1 +\const x^2)$. The horizon metric in
this case simplifies to \be \label{roundS3}\gamma_{ab}dx^a dx^b = \frac{\Gamma}{Q}
d\sigma^2 +(b_1+b_2 \sigma) (dx^1)^2 + \frac{2\sqrt{K}}{b_2
\Gamma^{1/2}} dx^1 dx^2 + \frac{ C^2(b_1-b_2
\sigma)+c_1b_2}{\Gamma b_2^2} (dx^2)^2 \; .\ee
We have now derived the most general local form of the metric in this class. It now remains to perform a global analysis of this solution by demanding that it extends to a regular metric on a compact 3d horizon. In fact, we may use a shortcut. Noting that the Ricci tensor of (\ref{roundS3}) is $R_{ab}= \frac{C^2}{2\Gamma} \gamma_{ab}$ we see that is must be locally isometric to the round metric on $S^3$. Therefore in this case $\mathcal{H}=S^3$ with a round metric.

Now consider the $b_2=0$ case which implies $b_1 \neq 0$ (otherwise $\gamma_{11} \equiv 0$). This gives
\be
\omega= \frac{C}{\Gamma^{1/2} b_1} \sigma
\ee
where again, without loss of generality, we have fixed the integration constant and overall sign. The shift freedom in $\sigma$ allows one to set $c_1=0$ so $Q=C^2(\sigma_2^2-\sigma^2)$. Then perform the change of coordinates $(\sigma,x^i) \to (\theta, \psi,\phi)$ defined by $\sigma=\sigma_2 \cos\theta$ and $x^2= \Gamma b_1^{1/2} \phi/(C^2\sigma_2)$ and $x^1= \Gamma^{1/2} \psi/(C b_1^{1/2})$, so $0 \leq \theta \leq \pi$ uniquely parameterises the interval. The horizon metric then becomes
\be
\gamma_{ab}dx^a dx^b= \frac{\Gamma}{C^2} \left[ d\theta^2+\sin^2\theta d\phi^2+ (d\psi+\cos\theta d\phi)^2 \right]
\ee
which is locally isometric to the round sphere on $S^3$.

Thus to summarise, all solutions in case (i) (i.e. $\Gamma$ a constant) corresponds to the round $S^3$ horizon.\\

\noindent {\bf Inhomogeneous $S^3$ horizon} \\

Here we consider case (ii)
with $\Gamma= \beta \sigma^2/4$. Substituting back into (\ref{k0B0eq}) gives $c_2=-4e^2/\beta$. We can integrate (\ref{gamma11ode}) to get
\be
\gamma_{11}= \frac{d_1}{\sigma}+d_2
\ee
 where $d_i$ are constants, and substituting this into (\ref{11eq}) gives
\be
\dot{\omega}^2= \frac{4K}{\beta^2(d_1+d_2\sigma)^4}
\ee
where $K \equiv d_1^2 \beta C^2+d_1d_2c_1\beta +4e^2 d_2^2 \geq 0$. Notice that equation (\ref{12eq}) is now automatically satisfied. To full determine the local form of the metric it remains to integrate for $\omega$; it turns out this depends on whether $d_2=0$ or not. In either case the metric is of the form
\be
\gamma_{ab}dx^a dx^b= \frac{\beta \sigma^2 d\sigma^2}{4 Q} + \gamma_{11}\left[ dx^1 + \omega dx^2 \right]^2 + \frac{4Q}{\beta \sigma^2 \gamma_{11}} (dx^2)^2
\ee
where as always $Q=-C^2 \sigma^2+c_1 \sigma+c_2$. We will also analyse regularity of these solutions -- we will find they all lead to smooth metrics on $S^3$ (or quotients). Recall that compactness implies $Q=C^2(\sigma-\sigma_1)(\sigma_2-\sigma)$ and $\sigma_1\leq \sigma \leq \sigma_2$. Also, since $\Gamma=\beta \sigma^2/4$ without loss of generality we can take $\sigma>0$, so $0<\sigma_1<\sigma_2$. Before moving on we note the identity $K= C^2 \beta (d_1+d_2\sigma_1)(d_1+d_2\sigma_2)$ which will be useful when performing our global analysis of the solutions.

First consider $d_2 \neq 0$. We can integrate to get
\be
\omega = \frac{2 \sqrt{K}}{\beta d_2( d_1 +d_2 \sigma)} \
\ee
where we have set the integration constant to zero and fixed the sign using the shift freedom $x^1 \to \pm( x^2 +\const x^1)$. There are a number of cases to consider depending on $d_1$.
\begin{itemize}
\item If $d_1>0$ then $\gamma_{11}$ is a monotonically decreasing function of $\sigma$. It follows that $\gamma_{11}(\sigma_2) \geq 0$ and $\gamma_{11}>0$ for $\sigma_1 \leq \sigma <\sigma_2$. There are thus two possibilities: either (i) $\gamma_{11}(\sigma_2)=0$ or (ii) $\gamma_{11}(\sigma_2)>0$. One can check the following useful identity $K= C^2 \beta (d_1+d_2\sigma_1)(d_1+d_2\sigma_2)$ which shows that in case (i) we have $K=0$ and in case (ii) $K>0$.

In case (i) we have $d_1+d_2\sigma_2=0$ and hence $\gamma_{11}= d_2(\sigma-\sigma_2)/\sigma$ so $d_2<0$. The horizon metric then reads
\be
\label{case1}
\gamma_{ab}dx^a dx^b = \frac{\beta \sigma^2 d\sigma^2}{4C^2(\sigma_2-\sigma)(\sigma-\sigma_1)}+ \frac{|d_2| (\sigma_2-\sigma)}{\sigma} (dx^1)^2+ \frac{4C^2(\sigma-\sigma_1)}{\beta |d_2| \sigma} (dx^2)^2
\ee
which is non-degenerate for $\sigma_1<\sigma<\sigma_2$. This metric has conical singularities at $\sigma=\sigma_1$ where $\partial_{x^2}$ vanishes and at $\sigma=\sigma_2$ where $\partial_{x^1}$ vanishes. Hence by periodically identifying $x^i$ appropriately one can remove these singularities leaving a smooth metric on $S^3$ (or quotients).

Now consider case (ii) for which we must have $\gamma_{11}>0$ for $\sigma_1 \leq \sigma \leq \sigma_2$ and $\omega(\sigma_1) \neq \omega(\sigma_2)$. It follows that the metric is non-degenerate for $\sigma_1 < \sigma <\sigma_2$ with conical singularities at $\sigma=\sigma_i$ where the killing field $\partial_{x^2}-\omega(\sigma_i) \partial_{x^1}$ vanishes. Further since these two Killing fields are distinct one can always remove these conical singularities leaving one with a smooth metric on $S^3$ (or quotients).

In fact case (ii) is isometric to case (i). To see this introduce coordinates $\phi^i$ so $m_i=\partial /\partial \phi^i= M_i\left(\partial_{x^2}-\omega(\sigma_i) \partial_{x^1}\right)$ where $M_i$ are constants chosen so $\phi^i$ have periods $2\pi$. It is then easy to check that the Killing part of the horizon metric in case (ii) is
\be
\label{killinggamma}
\gamma_{ij}dx^idx^j= \frac{4C^2 (\sigma_2-\sigma_1)}{\beta \sigma} \left[ \frac{M_1^2(\sigma-\sigma_1) (d\phi^1)^2}{(d_1+d_2\sigma_1)} + \frac{M_2^2(\sigma_2-\sigma) (d\phi^2)^2}{(d_1+d_2\sigma_2)}\right]
\ee
which allows one to see explicitly that the corresponding horizon metric is isometric to (\ref{case1}) (the constants in the Killing part of the metric are fixed in both cases to be the same by demanding the absence of conical singularities).
\item If $d_1=0$ then $\gamma_{11}=d_2$ and hence $d_2>0$. The metric is then
\be
\gamma_{ab}dx^a dx^b= \frac{\beta \sigma^2 d\sigma^2}{4 C^2(\sigma_2-\sigma)(\sigma-\sigma_1)} + d_2 \left[ dx^1 + \frac{4e}{\beta d_2 \sigma} dx^2 \right]^2 + \frac{4C^2(\sigma_2-\sigma)(\sigma-\sigma_1)}{\beta d_2\sigma^2} (dx^2)^2
\ee
which is non-degenerate for $\sigma_1<\sigma <\sigma_2$. At the endpoints $\sigma=\sigma_i$ the metric degenerates in such a way the Killing field $\partial_{x^2}- \omega(\sigma_i) \partial_{x^1}$ vanishes. Since $\omega(\sigma_1) \neq \omega(\sigma_2)$ it is a different Killing field which vanishes at each root. Hence one can always remove the corresponding conical singularities leading to a smooth metric on $S^3$ (or quotients). Introducing coordinates adapted to the rotational Killing fields $m_i$ as above one can show that the Killing part of the metric in this case is given by (\ref{killinggamma}) with $d_1=0$. Hence this case is also isometric to (\ref{case1}).

\item If $d_1<0$ then $\gamma_{11}$ is a monotonically increasing function of $\sigma$. It follows that $\gamma_{11}(\sigma_1) \geq 0$ and $\gamma_{11}>0$ for $\sigma_1 < \sigma \leq \sigma_2$. The analysis in this case is analogous to the $d_1>0$ case and again corresponds to smooth metrics on $S^3$ (or quotients). In fact the solutions in this class are isometric to the ones in the $d_1>0$ class.
\end{itemize}

It now remains to consider the $d_2=0$ case where one has
\be
\omega = \frac{2 \sqrt{K}}{\beta d_1^2} \sigma
\ee
again, without loss of generality, fixing the sign and setting the integration constant to zero. The metric is then
\be
\gamma_{ab}dx^adx^b = \frac{\beta \sigma^2 d\sigma^2}{4C^2(\sigma_2-\sigma)(\sigma-\sigma_1)}+ \frac{d_1}{\sigma} \left( dx^1+ \frac{2C}{\beta^{1/2} d_1} \sigma dx^2 \right)^2 + \frac{ 4C^2(\sigma_2-\sigma)(\sigma-\sigma_1)}{\beta \sigma d_1} (dx^2)^2
\ee
which again is non-degenerate for $\sigma_1<\sigma <\sigma_2$ and has conical singularities at the endpoints $\sigma=\sigma_i$ where $\partial_{x^2}-\omega(\sigma_i)\partial_{x^1}$ vanishes. Since $\omega(\sigma_1) \neq \omega(\sigma_2)$ the Killing fields are distinct has hence the conical singularties can be removed. Once more this leads to a regular metric on $S^3$ (or quotients). Introducing coordinates adapted to the rotational symmetries as above shows the Killing part of the metric is given by (\ref{killinggamma}) with $d_2=0$. Hence this case is also isometric to (\ref{case1}).

Therefore we have shown that all cases in the $\Gamma=\beta \sigma^2/4$ class are isometric to (\ref{case1}) (or equivalently (\ref{killinggamma})). Removing conical singularities leads to a smooth inhomogeneous metric on $S^3$:
\be
\gamma_{ab}dx^a dx^b = \frac{\beta}{C^2} \left[ \frac{\sigma^2 d\sigma^2}{4(\sigma-\sigma_1)(\sigma_2-\sigma)} + \frac{(\sigma-\sigma_1)\sigma_1^3 (d\phi^1)^2}{(\sigma_2-\sigma_1) \sigma} + \frac{(\sigma_2-\sigma)\sigma_2^3 (d\phi^2)^2}{(\sigma_2-\sigma_1) \sigma}  \right]
\ee
with $\phi^i \sim \phi^i+2\pi$ and $\sigma_1 \leq \sigma \leq \sigma_2$. We can parameterise $\sigma$ uniquely by $\cos^2\theta = (\sigma-\sigma_1)/(\sigma_2-\sigma_1)$ for $0 \leq \theta \leq \pi/2$, in terms of which the horizon metric is
\be\label{inhomS3}
\gamma_{ab}dx^a dx^b = \frac{\beta}{C^2} \left[ \sigma(\theta)^2 d\theta^2+ \frac{\sigma_1^3 \cos^2\theta}{\sigma(\theta)} (d\phi^1)^2+ \frac{\sigma_2^3 \sin^2\theta}{\sigma(\theta)} (d\phi^2)^2 \right]
\ee
where $\sigma(\theta)= \sigma_1\sin^2\theta + \sigma_2\cos^2\theta$.\footnote{Notice that in these coordinates one may set $\sigma_1=\sigma_2$ which gives the round metric on $S^3$.} The corresponding near-horizon solution is parameterised by the constants $(C^2,\beta,\sigma_i)$ with $e^2=C^2\beta\sigma_1\sigma_2/4$. However, the solution is really only a two parameter family. To see this, note that the the full near-horizon geometry is invariant under the following two independent scalings:
\begin{eqnarray}
\mathcal{S}_1&:& C^2\to LC^2, \qquad \beta \to L^{-1}\beta, \qquad \sigma_i \to  L\sigma_i, \qquad v \to L^{-1} v \label{S1}\\
\mathcal{S}_2&:& \beta \to \Omega^{-2}\beta, \qquad \sigma_i \to \Omega \sigma_i, \label{S2}
\end{eqnarray} where $(L,\Omega) > 0$. In particular note that under $\mathcal{S}_1$, $\Gamma \to L\Gamma, e\to Le$. As a consequence of these scaling symmetries, we may fix two combinations of the set $(C^2, \beta, \sigma_i)$ thus leaving two parameters.

In fact as shown in the Appendix, this near-horizon geometry corresponds to the near-horizon geometry of an extremal Reissner-Nordstr\"om black hole in a background electric field~\cite{EMDring}. Note that such regular solutions cannot occur in four dimensions as a consequence of the classification work of~\cite{CT}; indeed, the analogous solution suffers from a conical singularity in the extremal limit.

\subsubsection{No electric field or $\xi=0$}
This corresponds to $e=0$ or $\xi=0$. In this case we can solve (\ref{Beq}) for $B$ to get
\be
\star_2 B= \frac{p_i}{\Gamma \sqrt{\gamma}}dx^i
\ee
where $p_i$ are constants. It is convenient to exploit the freedom in choice of the coordinates $x^i$ so that $\star_2 B=\Gamma^{-1}\gamma^{-1/2}p dx^2$ where $p$ is a non-zero constant.
It follows that
\be
B_1= -\frac{p \gamma_{11}}{Q}, \qquad B_2=\omega B_1, \qquad B_iB^i= \frac{p^2 \gamma_{11}}{Q^2} \; .
\ee
This allows us to solve (\ref{A0}) for $\gamma_{11}$:
\be
\label{gamma11static} \gamma_{11} = \frac{Q}{p^2 \Gamma} \left[ \Gamma\frac{d}{d\sigma}\left( \frac{Q \dot{\Gamma}}{\Gamma} \right) +2C^2\Gamma-2e^2 \right] \; .
\ee
Also note that equation (\ref{gamma11ode}) in this case can be written as
\be
\label{gamma11odep2}
Q^2 \left[ \frac{d}{d\sigma} (\Gamma
\dot{\gamma}_{11} ) + \left( 2\ddot{\Gamma}-
\frac{\dot{\Gamma}^2}{\Gamma} \right)\gamma_{11} \right] + 3p^2 \gamma_{11}^2 \Gamma =0 \; ,
\ee
and equation (\ref{12eq}) tells us that $\dot{\omega}=k/(\gamma_{11}^2\Gamma)$ for constant $k$.
Now, we see that substituting our expression for $\gamma_{11}$ into (\ref{gamma11odep2}) gives a fourth order non-linear ODE for $\Gamma$. Given a solution to this ODE we see that $\gamma_{11}$ is determined and therefore $\omega$ can be solved for. This then determines the full near-horizon geometry. We have not been able to solve the 4th order ODE for $\Gamma$ in general, and as a result can not present a general local form for the metric in this class. Nevertheless, we can state the following: \\

\noindent {\bf Lemma 5}: There are no solutions such that $\Gamma \sim a \sigma^{n+2}$ as $\sigma \to \infty$, with $n>0$. Note that this implies that if $\Gamma$ is a polynomial its order must be at most quadratic.\\

\noindent {\bf Proof}: Assuming $\Gamma \sim a \sigma^{n+2}$, equation (\ref{gamma11static}) implies $\gamma_{11} \sim nC^4 \sigma^2 p^{-2}$. Equation (\ref{gamma11odep2}) in the limit $\sigma \to \infty$ then implies that $n(n+1)(n+6)=0$, which is a contradiction.\\

 Despite not being able to solve for the explicit metric in this
case, we will now show that global considerations allow one to
determine certain properties of solutions in this class including
the horizon topology. This is achieved by noticing that we can write
\be \label{gamma11inv} \gamma_{11} = \frac{1}{p^2} S\left(
\frac{S(\Gamma)}{\Gamma} \right) + \frac{Q}{p^2 \Gamma}(
2C^2\Gamma-2e^2 ) \ee where $S=Q \partial /\partial \sigma$ is a
globally defined vector field on $\mathcal{H}$ which vanishes at the
roots of $Q$. This expresses $\gamma_{11}$ in terms of globally
defined quantities on $\mathcal{H}$ and allows one to deduce that
$\gamma_{11}$ must vanish at both roots of $Q$. As we will now show, this class of
solutions {\it must} have $S^1 \times S^2$ topology horizons with
$\partial /\partial x^1$ being the Killing vector which vanishes at
the poles of the $S^2$.

Writing
$Q=C^2(\sigma_2-\sigma)(\sigma-\sigma_1)$ we see that
$\gamma_{11}=C_i(\sigma-\sigma_i)+ \mathcal{O}[(\sigma-\sigma_i)^2]$ for $\sigma
\to \sigma_i$ where $C_i$ are constants (note we must have $C_1>0$
and $C_2<0$)\footnote{If either $C_i=0$ then the metric will have
curvature singularities.}. It follows that $\dot{\omega} \sim
k\Gamma(\sigma_i)^{-1}C_i^{-2}(\sigma-\sigma_i)^{-2}$ near each root
and hence $\omega \sim
-k\Gamma(\sigma_i)^{-1}C_i^{-2}(\sigma-\sigma_i)^{-1}$. However
regularity demands that $\gamma_{12}=0$ at $\sigma=\sigma_i$ and
therefore we must have $k=0$. It follows that $\omega$ is a constant
which may be set to zero using the remaining freedom in the choice
of $x^i$ (i.e. shifting $x^1 \to x^1 +\const x^2$). The horizon
metric thus reads \be ds^2= \frac{\Gamma}{Q} d\sigma^2 +
\gamma_{11}(dx^1)^2 + \frac{Q}{\Gamma \gamma_{11}} (dx^2)^2 \ee
and we see that $\gamma_{22}>0$ for $\sigma_1 \leq \sigma \leq \sigma_2$. This metric is non-degenerate for $\sigma_1<\sigma<\sigma_2$ and has conical singularities at $\sigma=\sigma_1,\sigma_2$ in the $(\sigma,x^1)$ plane. Note that the Maxwell field also simplifies to
\be \hat{\cF}= -\frac{\sqrt{3}}{2} \frac{p\gamma_{11}}{Q}
d\sigma \wedge dx^1 \ee which is regular. We must now analyse
regularity of the horizon metric. The simultaneous removal of the
conical singularities at $\sigma=\sigma_i$ is equivalent to
\be
\label{reg}
C_1\Gamma_1^{-1}=-C_2 \Gamma_2^{-1}
\ee
where the constants
$C_i$ may be calculated by expanding (\ref{gamma11inv}) near
$\sigma=\sigma_i$: \be C_i = \frac{\dot{Q}_i^2 \dot{\Gamma}_i}{p^2 \Gamma_i} + \frac{2 \dot{Q}_i}{p^2 \Gamma_i} (C^2 \Gamma_i-e^2)\; . \ee If the condition (\ref{reg}) is met then the
horizon metric is a smooth metric on $S^1 \times S^2$ which
generically is inhomogeneous (with $x^2$ a
coordinate on $S^1$). The corresponding near-horizon
geometry would then be a static geometry with black ring like
topology. Note that one might suspect that the other equations of motion could constrain the boundary conditions further -- in fact this is not the case. We find that equations (\ref{gamma11odep2}) and (\ref{11eq}) expanded near $\sigma=\sigma_i$ give simultaneous equations for $C_i$ and $D_i$, where $\gamma_{11}=C_i(\sigma-\sigma_i)+D_i(\sigma-\sigma_i)^2/2 +\dots$, which can be solved to give the same value of $C_i$ as we obtained above. These results may be summarised by the following:\\

\noindent {\bf Lemma 6:} Consider an AdS$_2$ near-horizon geometry with no electric field for $\xi \neq 0$, or an AdS$_2$ near-horizon geometry for $\xi=0$. Assume that the solution has a further $U(1)^2$ symmetry with closed space-like orbits and that spatial sections of the horizon, $\mathcal{H}$, are compact. Then $\mathcal{H}$ must have $S^1 \times S^2$ topology. \\

The above discussion does not, of course, address the crucial question of existence of solutions, which we now turn to.

First let us consider solutions with $\Gamma$ a constant which, from equation (\ref{gamma11inv}), must have
\be
\gamma_{11}=  \frac{KQ}{\Gamma}, \qquad K \equiv \frac{2(C^2\Gamma-e^2)}{p^2}
\ee
where the constant $K>0$. Substituting into (\ref{gamma11odep2}) implies $e^2=\frac{2}{3}C^2\Gamma$. Therefore we must have $e \neq 0$ and hence this solution is only valid for $\xi=0$ (i.e. in pure Einstein-Maxwell theory). Thus we see that for $\xi \neq 0$, which includes minimal supergravity, there are no $\Gamma$ constant solutions in this class. The horizon metric and Maxwell field for the $\xi=0$ solution are
\be
\gamma_{ab} = \frac{\Gamma d\sigma^2 }{Q} + \frac{KQ}{\Gamma} (dx^1)^2+K(dx^2)^2 \qquad
\hat{\cF} = -\frac{\sqrt{3}}{2} pK\Gamma^{-1} d\sigma \wedge dx^1 \; ,
\ee
which are regular on $S^2\times S^1$ provided the period of $x^1$ is chosen appropriately. We can bring this solution into a simpler form by translating $\sigma$ so as to set $c_1=0$ and introducing new coordinates $\theta, \phi$ as follows: $\sigma /\sigma_2 =\cos \theta$, with $0 \leq \theta \leq \pi$; $\phi= C^2 \sigma_2 \sqrt{K} x^1/\Gamma$; $z=\sqrt{\frac{2C^2\Gamma}{3p^2}} x^2$. The full near-horizon geometry is then
\bea
\label{ads2s2s1}
ds^2 &=& \frac{3e^2}{2C^2}\left[ -C^2 r^2 dv^2 +2dvdr + \frac{1}{C^2} \left( d\theta^2 +\sin^2\theta d\phi^2 \right) \right] + dz^2, \\
\cF &=& \frac{\sqrt{3}}{{2}} \left[ e dr \wedge dv + \frac{|e| |p| }{pC^2} \sin\theta d\theta \wedge d\phi \right]
\eea
which is simply the direct product $AdS_2 \times S^2 \times S^1$. In fact the radii of the $AdS_2$ and $S^2$ are equal and given by $\ell^2 \equiv C^2/\Gamma= 2C^4/3e^2$. Notice the parameter $p$ only appears in the combination $|p|/p=\textrm{sgn}(p)$. Thus there are two solutions (one for each $\textrm{sgn}(p)$) parameterised by $(C^2,e, \Delta z)$, although due to a scaling symmetry only $(\ell, \Delta z)$ are non-trivial. We therefore have a two two-parameter families labelled by the radii of the $S^2$ and $S^1$ with the Maxwell field determined in terms of these. In fact these two solutions correspond to the near-horizon geometry of certain dyonic strings. These are constructed by taking the direct product of the dyonic $Q=\pm P$ extremal Reissner-Nordstr\"om solution of 4d Einstein-Maxwell with an $S^1$. Note that in general one cannot KK reduce solutions to 5d Einstein-Maxwell to 4d Einstein-Maxwell -- in this special case though it works because the solution is static and $\cF^2=0$.

Now consider $\Gamma$ nonconstant solutions. In all known extremal black hole examples (in $D=4,5$ with any topology and asymptotics) $\Gamma$ is a quadratic polynomial in $\sigma$. It is easy to check that for solutions of the form $\Gamma=a_0+a_1\sigma$ the regularity condition $C_1\Gamma_1^{-1}=-C_2 \Gamma_2^{-1}$ implies $\Gamma_1=\Gamma_2$ which can never be satified since $\sigma_2>\sigma_1$. For $\Gamma$ quadratic, by shifting $\sigma$, without loss of generality we can set the linear term to zero, so $\Gamma=a_0+a_2\sigma^2$. One can then check that the regularity condition again implies $\Gamma_1=\Gamma_2$. This then implies that $\sigma_2=-\sigma_1$ and thus $c_1=0$.\footnote{In fact note the regularity condition can be satisfied if $c_1=0$ and $\Gamma$ is an even function of $\sigma$.} Let us now consider the $\Gamma=a_0+a_2\sigma^2$ case further. It can be shown that the fourth order ODE for $\Gamma$ implies $a_0C^2+a_2c_2=0$ in both the $e=0$ and $(e \neq 0, \xi=0)$  cases. However this constraint on the parameters can also be written as $\Gamma(\sigma_2)=0$ which contradicts the fact that $\Gamma>0$. This rules out the existence of regular near-horizon geometries of this kind with $\Gamma$ a nonconstant quadratic polynomial.

Together with our Lemma 5 above, these results provide some evidence for the following possibilities (which we have not proved): (i) regular static $AdS_2$ near-horizon geometries with compact horizon, $U(1)^2$ symmetry, and vanishing electric field, for any Chern-Simons coupling $\xi$, do not exist; (ii) the only regular static $AdS_2$ near-horizon geometry with a compact horizon, $U(1)^2$ symmetry, and non-zero magnetic field in Einstein-Maxwell (i.e. $\xi=0$) is the direct product solution $AdS_2 \times S^2 \times S^1$ given by (\ref{ads2s2s1}).

\subsubsection{Non-zero electric and magnetic fields and $\xi \neq 0$}

In this section we will analyse the remaining case $\xi \neq 0$ with
$e \neq 0$ and $B \neq 0$. It turns out this is the most non-trivial
possibility for static near-horizon geometries. The remaining part
of the Maxwell equation (\ref{Beq}) is equivalent to two first order
equations. Defining $G\equiv \gamma_{11}\Gamma(B_2-\omega
B_1)$ and $H \equiv B_1$ one can show
that (\ref{Beq}) is equivalent to \be \label{GHodes} \dot{G}+2\xi e
H=0, \qquad -\Gamma \frac{d}{d\sigma}\left(\frac{Q H}{\gamma_{11}}
\right)+ G \left( \Gamma \dot{\omega} + \frac{2\xi e}{\gamma_{11}}
\right) =0 \; .\ee
Note that in terms of these variables
\be
\hat{\cF}= \frac{\sqrt{3}}{2} \left[ \frac{G}{\Gamma \gamma_{11}} d\sigma \wedge dx^2+ H d\sigma \wedge (dx^1+\omega dx^2) \right] \; .
\ee
Since we are assuming $\xi e \neq 0$ these ODEs
are coupled and the system of equations is also equivalent to the
single second order ODE: \be \label{Gode} \Gamma
\frac{d}{d\sigma}\left(\frac{Q \dot{G}}{\gamma_{11}} \right)+ 2\xi
eG \left( \Gamma \dot{\omega} + \frac{2\xi e}{\gamma_{11}} \right)
=0 \; .\ee It is also worth noting that in terms of these variables
\be B_iB^i= \frac{Q\Gamma H^2+G^2}{Q\Gamma \gamma_{11}} \ee  from
which it follows that $B_iB^i \equiv 0$ if and only if $G\equiv 0$
-- thus in this section we are assuming $G \neq 0$. Next note that
equation (\ref{12eq}) is \be \frac{d}{d\sigma}( \Gamma \gamma_{11}^2
\dot{\omega} ) +3GH=0  \ee which can then be integrated with the
help of (\ref{GHodes}) giving \be \label{12Geq} \dot{\omega}= \frac{3G^2+k}{4\xi
e\Gamma\gamma_{11}^2} \ee for some constant $k$. This can be used to
eliminate $\dot{\omega}$ in (\ref{Gode}) which then reads \be
\label{Godek}\Gamma \gamma_{11}^2\frac{d}{d\sigma}\left(\frac{Q
\dot{G}}{\gamma_{11}} \right)+ G\left( \frac{3G^2+k}{2} +4\xi^2 e^2
\gamma_{11} \right)=0 \; .\ee Now, observe that (\ref{A0}) can be
written as \be \label{gamma11G} \gamma_{11}\left(
\Gamma\frac{d}{d\sigma}\left( \frac{Q \dot{\Gamma}}{\Gamma} \right)
+2C^2\Gamma-2e^2 \right)= \frac{Q\Gamma \dot{G}^2}{4\xi^2e^2} +G^2
\ee which can be solved for $\gamma_{11}$ as a function of
$(\Gamma,G)$ wherever $B_iB^i \neq 0$ (since by equation
(\ref{A0}) the factor on the LHS is then non-zero). One can then
substitute this expression for $\gamma_{11}$ into (\ref{Godek}) to
get a non-linear ODE for $(\Gamma,G)$. Another such ODE can be
derived by substituting the expression for $\gamma_{11}$ into \be
\label{gamma11odeG} \frac{d}{d\sigma} (\Gamma \dot{\gamma}_{11} ) +
\left( 2\ddot{\Gamma}- \frac{\dot{\Gamma}^2}{\Gamma}
\right)\gamma_{11} + \frac{3}{4\xi^2 e^2} \Gamma \dot{G}^2 =0 \; \ee
which comes from (\ref{gamma11ode}). We have thus reduced the
problem to solving two coupled non-linear ODEs for $(\Gamma,G)$. We
now turn to finding specific solutions.

Notice that so far we have not used the assumption of compactness of $\mathcal{H}$. As we proved in our theorem earlier this turns out to be particularly restrictive for this class of near-horizon geometries -- it implies that $\Gamma$ must be a constant. We will examine the near-horizon equation under this assumption now. First, note that for $G \neq 0$ (\ref{gamma11G}) implies the constant $C^2\Gamma-e^2>0$ and we can solve for $\gamma_{11}$:
\be
\gamma_{11}= \frac{Q\Gamma \dot{G}^2 +4\xi^2e^2G^2}{8\xi^2e^2(C^2\Gamma-e^2)}
\ee
and this can be substituted into (\ref{gamma11odeG}) to give
\be
\label{G3ode}
\frac{d^2}{d\sigma^2} \left[ Q\Gamma \dot{G}^2 +4\xi^2e^2G^2 \right] +6(C^2\Gamma-e^2) \dot{G}^2=0
\ee
which is a third order non-linear ODE for the function $G$. The classification of near-horizon geometries in this class thus reduces to solving this ODE. Clearly $\dot{G}=0$ is a solution to this equation. A more general solution which includes this is given by $G=g_0+g_1\sigma$ for constants $g_i$. Note that if $g_1 \neq 0$ we can always use the shift freedom in the definition to $\sigma$ to set $G=g \sigma$. We will find it convenient to thus analyse $G$ constant and $G$ linear separately.

\paragraph{$G$ constant} In this case we assume $G$ is a non-zero
constant and thus $H=0$. Then our expression for
$\gamma_{11}$  gives $\gamma_{11}=
G^2/(2C^2\Gamma-2e^2)$. Also note that (\ref{Gode}) implies
$\dot{\omega}=- 2\xi e/(\Gamma \gamma_{11})$. However, (\ref{11eq})
gives us another equation $\gamma_{11}^2 \Gamma^2\dot{\omega}^2=
e^2-G^2/\gamma_{11}$. Eliminating $\dot{\omega}$ implies
$1-4\xi^2>0$ and \be \gamma_{11}=\frac{G^2}{e^2(1-4\xi^2)} \ee which
upon comparing to our first expression for $\gamma_{11}$ gives \be
C^2\Gamma= \frac{(3-4\xi^2) e^2}{2} \; .\ee We can integrate for
$\omega$ to get $\omega =\omega_0 \sigma + \const$ where
$\omega_0=-2\xi e/(\Gamma \gamma_{11})$ is a constant which works
out to be \be \omega_0= -\frac{4C^2 \xi e(1-4\xi^2)}{G^2(3-4\xi^2)}
\; . \ee We have thus fully determined the form of the near-horizon
data which reads (shifting $x^1$ so as to set the integration
constant in $\omega$ to zero) \bea \gamma_{ab}dx^a dx^b &=&
\frac{\Gamma}{Q} d\sigma^2+ \gamma_{11}( dx^1+ \omega_0 \sigma
dx^2)^2 + \frac{Q}{\Gamma \gamma_{11}} (dx^2)^2 \\ \hat{\cF} &=&
\frac{\sqrt{3}}{2} \frac{G}{\Gamma \gamma_{11}} d\sigma \wedge dx^2 \eea
where $Q=-C^2\sigma^2+c_1 \sigma +c_2$ and
$\gamma_{11},\Gamma,\omega_0$ are constants as above all determined
in terms of the constants $G,e$. This metric is locally isometric to a regular
homogeneous metric on $S^3$ as we will show next. First though, note that its derivation shows it is only valid for
$0<\xi^2<1/4$ and thus does not exist in supergravity for which
$\xi=1$. However one can take its $\xi \to 0$ limit to obtain the
$AdS_2 \times S^1 \times S^2$ solution we derived in the previous
section.

We may used the translation freedom in defining $\sigma$ to set $c_1=0$. Then $Q=C^2(\sigma_2^2-\sigma^2)$ where $\sigma_2^2=c_2/C^2$. Now change coordinates to
\be
\cos\theta = \frac{\sigma}{\sigma_2}, \qquad \psi= \frac{C^2}{\Gamma\omega_0 \sqrt{\gamma_{11}}}x^1, \qquad \phi= \frac{C^2 \sigma_2}{\Gamma \sqrt{\gamma_{11}}}x^2
\ee
so $0\leq \theta \leq \pi$ uniquely parameterises the interval $[-\sigma_2,\sigma_2]$. The near-horizon solution then becomes
\bea
&&ds^2= \frac{(3-4\xi^2) e^2}{2C^2}\left[ -C^2r^2 dv^2 +2dvdr+ \frac{1}{C^2}(d\theta^2+\sin^2\theta d\phi^2) \right]+ \frac{4\xi^2 e^2}{C^4}( d\psi+\cos\theta d\phi)^2 \nonumber \\
&&\cF= \frac{\sqrt{3}}{2} \left[ e dr \wedge dv+  \frac{|G||e| \sqrt{1-4\xi^2}}{G C^2} d\cos\theta \wedge d\phi \right]
\eea
so the geometry consists of a direct product of AdS$_2$ and a homogeneously squashed $S^3$ (if $0 \leq \phi \leq 2\pi$ and $0 \leq \psi \leq 4\pi$) as claimed. Notice that all dependence of the constant $G$ has cancelled from the metric and appears only in the combination $|G|/G=\textrm{sgn}(G)$ in the Maxwell field. This thus describes two one-parameter families labelled by $e, \textrm{sgn}(G)$ ($C^2$ is a trivial parameter). Notice that this is the same solution we derived in Lemma 2 earlier -- indeed one can check that for this solution $\cA =\const \partial/ \partial \psi$ is Killing and $\cA^2=3(1-4\xi^2)/(16\xi^2)$.

\paragraph{$G$ nonconstant} It is easy to check that $G=g \sigma$ is a solution to (\ref{G3ode}) provided $C^2\Gamma = \frac{e^2(3-4\xi^2)}{2}$ is satisfied. Notice this is the same as in the $G$ constant case. Also it implies that $C^2\Gamma-e^2=\frac{e^2(1-4\xi^2)}{2}$ and hence $1-4\xi^2>0$ in this case too. It follows that
\be\label{g11Gnonconst}
\gamma_{11}= \frac{g^2( Q\Gamma+4\xi^2e^2 \sigma^2)}{4\xi^2 e^4(1-4\xi^2)}
\ee
and from (\ref{Gode})
\be
\dot{\omega}= \frac{8\xi^3 e^5(1-4\xi^2)[ 3C^2\sigma^2(4\xi^2-1)+c_2(4\xi^2-3)]}{g^2(4\xi^2-3)( \Gamma Q +4\xi^2 e^2 \sigma^2)^2}
\ee
This can be integrated to give:
\be
\omega= \omega_0+\frac{16C^2 e^3 \xi^3 (1-4\xi^2) \sigma}{(3-4\xi^2)g^2 (\Gamma Q +4\xi^2 e^2 \sigma^2)}
\ee
where $\omega_0$ is an integration constant. One can also check that (\ref{12Geq}) is satisfied with the constant $k=\frac{g^2c_2(4\xi^2-3)}{C^2(4\xi^2-1)}$. One can now check that the remaining near-horizon equations are satisfied (it suffices to check (\ref{11eq})). The horizon metric is given by (\ref{canform}). The Maxwell field simplifies to
\be
\hat{\cF}= -\frac{\sqrt{3}}{2} \frac{g}{2\xi e} d\sigma \wedge (dx^1+\omega_0 dx^2) \; .
\ee
We now consider regularity of this solution. Recall that compactness requires $Q=C^2(\sigma_2-\sigma)(\sigma-\sigma_1)$ with $\sigma_1<\sigma_2$ and $\sigma_1 \leq \sigma \leq \sigma_2$. The analysis splits into two cases: either both roots $\sigma_i$ are non-zero or one of them vanishes (both cannot vanish as they must be distinct). Consider the case where both roots are non-zero. From the form of the solution we see that $\gamma_{11}>0$ for $\sigma_1 \leq \sigma \leq \sigma_2$. Therefore the metric on $\mathcal{H}$ is positive definite and non-degenerate for $\sigma_1< \sigma < \sigma_2$ and degenerates at $\sigma=\sigma_i$. The vector field $m_i= d_i( \partial/\partial x^2 -\omega(\sigma_i) \partial /\partial x^1)$ vanishes at $\sigma=\sigma_i$ where $d_i$ is constant and in general the metric possesses conical singularities at these points. Since $\omega(\sigma_1) \neq \omega(\sigma_2)$ it is a different vector field which vanishes at each root. By choosing the constants $d_i$ appropriately one may therefore remove these conical singularities. This leaves one with a smooth inhomogeneous metric on $S^3$ (or quotients). Defining coordinates adapted to the $m_i=\partial /\partial \phi^i$ we see that the $x^i$ are some linear combination of $\phi^i$; it is then easy to see that the horizon field strength $\hat{\cF}$ is regular everywhere too.

We now consider the second case in which one of the roots vanishes. Without loss of generality we take $\sigma_1 =0$ and $\sigma_2 > 0$ so that $Q = C^2(\sigma_2-\sigma)\sigma$.  From~(\ref{g11Gnonconst}) we see $\gamma_{11}$ vanishes at $\sigma=0$. The metric on~$\mathcal{H}$ hence is positive-definite and degenerate at $\sigma=0,\sigma_2$ where the distinct Killing fields $m_1 = d_1\partial/\partial x^1$ and $m_2 = d_2\left(\partial/\partial x^2 - \omega(\sigma_2)\partial/\partial x^1\right)$ vanish respectively. In general the metric will have conical singularities at these points, but it is always possible to choose the constants $d_i$ to ensure regularity, i.e. $m_i = \partial/\partial \phi^i$ where $\phi^i$ have period $2\pi$. Since it each of the distinct Killing fields has one fixed point, we have an inhomogeneous metric on $S^3$ (or quotients). The horizon field strength $\hat{\cF}$ can also be shown to be regular everywhere by writing it in terms of the $\phi^i$. In particular, near the degeneration points it is proportional to the associated to the volume form of the associated $\mathbb{R}^2$. \\

To summarise, the near-horizon geometries we have derived in this section are direct products of AdS$_2$ with either a homogeneous ($G$ constant) or an inhomogeneous ($G$ nonconstant) metric on $S^3$ and a Maxwell field with both electric and magnetic components turned on. Furthermore, they are only valid for CS coupling $0<\xi^2<1/4$. We have not been able to solve the ODE (\ref{G3ode}) in general and thus it is possible there are other $G$ nonconstant solutions. Thus we are not able present a complete classification of near-horizon geometries with both electric and magnetic fields turned on (and $\xi \neq 0$).

\section{Summary}
In this section we summarise the main results we have gathered for static near-horizon geometry solutions (with compact horizon sections and $U(1)^2$ symmetry) to $D=5$ Einstein-Maxwell coupled to a Chern-Simons term with coupling $\xi$. First recall that there are two ways a 5d near-horizon geometry with $U(1)^2$ symmetry can be static: either a warped product of (a quotient of) AdS$_3$ with some compact $M_2$ or a warped product of AdS$_2$ with $\mathcal{H}$~\cite{KLR2}:

\paragraph{AdS$_3$} In this case we have proved the most general near-horizon geometry is given by the direct product of a quotient of a patch of AdS$_3$ with a round $S^2$ (spatial cross-sections of the horizon have $S^1 \times S^2$ geometry). Note this is a solution for any $\xi$ and thus is valid for both pure Einstein-Maxwell and minimal supergravity. In fact the near-horizon geometry of the asymptotically flat supersymmetric black ring~\cite{R} and the asymptotically KK supersymmetric black string~\cite{Bena} are both in this class\footnote{The near-horizon region of the static extremal non-BPS ring discussed in~\cite{Emp} is given by a quotient of a different patch of AdS$_3 \times S^2$ (the Poincare patch). However, this example is a null orbifold singularity rather than a regular black hole (the horizon corresponds to the Poincare horizon of AdS$_3$.). Such solutions are not covered by our analysis.}.

\paragraph{AdS$_2$} In this case we can classify all cases with electric but no magnetic
fields for any $\xi$ and find that there are two solutions both with
$S^3$ horizons: one is the simple direct product $AdS_2 \times S^3$
and the other is a warped product of $AdS_2$ with an inhomogeneous
metric on $S^3$. These correspond to the near-horizon limits of asymptotically flat extremal RN and extremal RN in a background electric field respectively. In the case of a magnetic but no electric field we reduce the problem to a single non-linear $4$th order ODE and could not find any explicit solutions to this, although we could show that any such solution must have $S^1 \times S^2$ horizon topology. When one has both electric and magnetic fields we reduce the problem to one third order non-linear ODE. We find no examples of this kind for $\xi^2 \geq  1/4$ (which includes the case of minimal supergravity), for $0<\xi^2<1/4$ we find two examples with $S^3$ horizon topology (one homogeneous and one inhomogeneous metric), whereas for $\xi=0$ the only example we have is the direct product $AdS_2 \times S^2 \times S^1$ (which is in fact the NH limit of a dyonic string constructed by oxidising extremal RN with $Q=\pm P$ to 5d). \\

\noindent These results should provide a starting point towards solving the classification problem for asymptotically flat static extremal black holes in five dimensional Einstein-Maxwell-CS theory. As we have shown, the presence of a magnetic field complicates even the classification of near-horizon geometries. Furthermore, even with an electric but no magnetic field we have shown there are two possible near-horizon geometries with the same topology: one corresponds to the round metric on $S^3$ and the other to an inhomogeneous metric on $S^3$. Note that the latter case was not found in~\cite{Rog2}. This suggests that extending the uniqueness theorem of~\cite{GI
SEM} to extremal black holes would require proving that the near-horizon geometry with an inhomogeneous horizon cannot be the near-horizon limit of an asymptotically flat black hole.\\

\noindent {\bf Acknowledgements}: HKK and JL are supported by STFC.

\appendix
\section{Extremal Reissner-Nordstr\"om in a Background Field}
In this appendix we give the near-horizon geometry of an extremal 5d Reissner-Nordstr\"om black hole in an external background electric field. To obtain this solution we will apply a solution-generating technique as in~\cite{EMDring} based on an analogue of the Harrison transformation~\cite{Gal}. The Reissner-Nordstr\"om black hole is given by
\be\label{RN}
ds^2=-Vdt^2+ \frac{dr^2}{V}+ r^2(d\theta^2+\sin^2\theta d\phi^2+\cos^2\theta d\psi^2), \qquad A=\frac{\sqrt{3}}{2}\frac{Q}{r^2}dt
\ee
where $V= 1-2M/r^2+Q^2/r^4$.  First dualise $H=\star F$ and calculate the potential $B$ defined by $H=dB$. One gets
\be
B= \frac{\sqrt{3}}{2} \frac{Q}{2} \left( \cos^2 \theta -\frac{1}{2} \right) d\psi \wedge d\phi
\ee
where we have defined the orientation by $\epsilon_{tr\theta\psi\phi}>0$. The solution generating procedure~\cite{Gal} can be applied to metrics of this kind with a three-form with just the $B_{\psi\phi}$ component switched on as is the case here.
The transformed solution is
\bea
\label{extRNfull}
ds^2 &=& P^2\left[ -Vdt^2+ \frac{dr^2}{V}+ r^2d\theta^2 \right]+ \frac{r^2}{P}( \sin^2\theta d\phi^2+\cos^2\theta d\psi^2) \\
B &=& \frac{\sqrt{3}}{2} \frac{Q}{2 P} \left( \cos^2\theta -\frac{1}{2}-\frac{cQ}{4} \right) d\psi \wedge d\phi
\eea
where $c$ is the constant appearing in the generating procedure (so $c=0$ reduces to RN) and
\be
P=\left( 1- \frac{cQ}{2} \cos 2\theta \right)^2+c^2r^4\sin^2\theta \cos^2\theta
\ee
and regularity requires $P>0$ everywhere on and outside the horizon. This can be achieved by choosing $cQ<2$. It is a simple task to dualize back to find $F = -\star H$.  It can be checked that the metric~(\ref{extRNfull}) has a regular event horizon of topology $S^3$ at the largest root of $V$, although  the horizon is equipped with an inhomogeneous metric whereas the horizon of the seed solution~(\ref{RN}) is a round $S^3$. Hence~(\ref{extRNfull}) has the interpretation of being a Reissner-Nordstr\"om black hole immersed in a background electric fluxbrane~\cite{EMDring} which `distorts' the horizon.

Notice that one obtains an extremal black hole for $M=|Q|$ as for the asymptotically flat seed. Without loss of generality let $Q> 0$. Taking the near-horizon limit is straightforward and one finds the following solution parametrised by $(Q,c)$:\footnote{Note that the $r$ coordinate here is different to the original one in (\ref{extRNfull}).}
\begin{eqnarray}\label{extRN}
ds^2 &=& \frac{P(\theta)^2}{2Q^{1/2}}\left(-\frac{2r^2dv^2}{Q^{3/2}} + 2dvdr\right) + Q\left[P(\theta)^2d\theta^2 + \frac{P(0)^3\sin^2\theta(d\hat\phi)^2}{P(\theta)} + \frac{P(\frac{\pi}{2})^3\cos^2\theta(d\hat\psi)^2}{P(\theta)}\right] \nonumber \\
\cF &=& \frac{\sqrt{3}}{2}d( \alpha r dv), \qquad \alpha=\frac{1}{Q}\left(\frac{c^2Q^2}{4} - 1\right)
\end{eqnarray} and the hatted angles each have period $2\pi$ and
\begin{equation}
 P(\theta) = \left(1 + \frac{cQ}{2}\right)^2 - 2cQ\cos^2\theta.
\end{equation}

Now consider the near-horizon geometry with vanishing magnetic field derived previously in Section \ref{inhom}. The full five-dimensional metric is given by (we have sent $r\to \Gamma r$):
\begin{eqnarray}\label{inhomS3NH}
ds^2 &=& \Gamma(\hat\theta)\left(-C^2r^2dv^2 + 2dvdr\right) + \frac{\beta}{C^2} \left[ \sigma(\hat\theta)^2 d\hat\theta^2+ \frac{\sigma_1^3 \cos^2\hat\theta}{\sigma(\hat\theta)} (d\phi^1)^2+ \frac{\sigma_2^3 \sin^2\hat\theta}{\sigma(\hat\theta)} (d\phi^2)^2 \right] \\
\cF &=& \frac{\sqrt{3}}{2}d\left[e r dv\right], \qquad e^2 = \frac{C^2\sigma_1\sigma_2\beta}{4} \nonumber
\end{eqnarray}
where $\sigma(\hat\theta)= \sigma_1\sin^2\hat\theta + \sigma_2\cos^2\hat\theta$, $\Gamma(\hat\theta) = \beta \sigma(\hat\theta)^2 /4$, and $\phi^i$ have period $2\pi$. We will now show~(\ref{inhomS3NH}) is isometric to~(\ref{extRN}). Firstly define the constants
\begin{equation}
Q \equiv \frac{2}{\sqrt{C^2\beta}}, \qquad c \equiv \frac{(\sigma_1 - \sigma_2)\sqrt{C^2\beta}}{4}.
\end{equation} The first of these is invariant under the scaling symmetry~(\ref{S1}) and the second is invariant under~(\ref{S2}). Now use~(\ref{S1}) to set $C^2 = 2Q^{-3/2}$.  This choice then implies $\beta = 2 Q^{-1/2}$. Then use the second scaling symmetry~(\ref{S2}) to fix the following relation:
\begin{equation}\label{relation}
(\sigma_1 - \sigma_2)^2 + 16 = 8(\sigma_1 + \sigma_2).
\end{equation} This is possible because each term in the above equation transforms in a different way under~(\ref{S2}). Using $\sigma_2 > \sigma_1 > 0$ it is easy to check one can always choose the scaling parameter $\Omega$ defined in~(\ref{S2}) to ensure~(\ref{relation}) holds. Note that $C^2$ is not affected by this second scaling. With these choices we can invert to find
\begin{equation}
\sigma_{1} = \left(1+\frac{cQ}{2}\right)^2, \qquad \sigma_{2} = \left(1-\frac{cQ}{2}\right)^2.
\end{equation} Finally making the identifications
\begin{equation}
\theta = \hat\theta, \qquad \hat\phi = \phi^2, \qquad \hat\psi = \phi^1,
\end{equation} so that $\sigma(\hat\theta) = P(\theta)$, one can check that~(\ref{inhomS3NH}) is isometric to~(\ref{extRN}). One can also easily check that $e=\alpha$ confirming the Maxwell fields agree too.


\begin{thebibliography}{99}
{\small
\bibitem{GS}
  G.~J.~Galloway and R.~Schoen,
  ``A generalization of Hawking's black hole topology theorem to higher
  dimensions,''
  Commun.\ Math.\ Phys.\  {\bf 266} (2006) 571
  [arXiv:gr-qc/0509107].

\bibitem{Galloway}
  G.~J.~Galloway,
  ``Rigidity of outer horizons and the topology of black holes,''
  arXiv:gr-qc/0608118.

\bibitem{HIW}
  S.~Hollands, A.~Ishibashi and R.~M.~Wald,
  ``A Higher Dimensional Stationary Rotating Black Hole Must be Axisymmetric,''
  Commun.\ Math.\ Phys.\  {\bf 271} (2007) 699
  [arXiv:gr-qc/0605106].

\bibitem{HI}
  S.~Hollands and A.~Ishibashi,
  ``On the `Stationary Implies Axisymmetric' Theorem for Extremal Black Holes
  in Higher Dimensions,''
  arXiv:0809.2659 [gr-qc].

\bibitem{IM}
 V.~Moncrief and J.~Isenberg,
  ``Symmetries of Higher Dimensional Black Holes,''
  Class.\ Quant.\ Grav.\  {\bf 25} (2008) 195015
  [arXiv:0805.1451 [gr-qc]].

\bibitem{SW}
  D.~Sudarsky and R.~M.~Wald,
  ``Mass Formulas For Stationary Einstein Yang-Mills Black Holes And A Simple
  Proof Of Two Staticity Theorems,''
  Phys.\ Rev.\  D {\bf 47} (1993) 5209
  [arXiv:gr-qc/9305023].

\bibitem{GISvac}
  G.~W.~Gibbons, D.~Ida and T.~Shiromizu,
  ``Uniqueness and non-uniqueness of static black holes in higher
  dimensions,''
  Phys.\ Rev.\ Lett.\  {\bf 89}, 041101 (2002)
  [arXiv:hep-th/0206049].

\bibitem{GISEM}
  G.~W.~Gibbons, D.~Ida and T.~Shiromizu,
  ``Uniqueness of (dilatonic) charged black holes and black p-branes in  higher
  dimensions,''
  Phys.\ Rev.\  D {\bf 66}, 044010 (2002)
  [arXiv:hep-th/0206136].


\bibitem{CT}
  P.~T.~Chrusciel and P.~Tod,
  ``The classification of static electro-vacuum space-times containing an
  asymptotically flat spacelike hypersurface with compact interior,''
  Commun.\ Math.\ Phys.\  {\bf 271} (2007) 577
  [arXiv:gr-qc/0512043].

\bibitem{Rog1}
  M.~Rogatko,
  ``Uniqueness theorem of static degenerate and non-degenerate charged  black
  holes in higher dimensions,''
  Phys.\ Rev.\  D {\bf 67} (2003) 084025
  [arXiv:hep-th/0302091].

\bibitem{Rog2}
  M.~Rogatko,
  ``Classification of static charged black holes in higher dimensions,''
  Phys.\ Rev.\  D {\bf 73} (2006) 124027
  [arXiv:hep-th/0606116].


\bibitem{BMPV}
  J.~C.~Breckenridge, R.~C.~Myers, A.~W.~Peet and C.~Vafa,
  ``D-branes and spinning black holes,''
  Phys.\ Lett.\  B {\bf 391} (1997) 93
  [arXiv:hep-th/9602065].

\bibitem{susyring}
  H.~Elvang, R.~Emparan, D.~Mateos and H.~S.~Reall,
  ``A supersymmetric black ring,''
  Phys.\ Rev.\ Lett.\  {\bf 93}, 211302 (2004)
  [arXiv:hep-th/0407065]

\bibitem{R}
  H.~S.~Reall,
  ``Higher dimensional black holes and supersymmetry,''
  Phys.\ Rev.\  D {\bf 68} (2003) 024024
  [Erratum-ibid.\  D {\bf 70} (2004) 089902]
  [arXiv:hep-th/0211290].

\bibitem{KLvac}
  H.~K.~Kunduri and J.~Lucietti,
  ``A classification of near-horizon geometries of extremal vacuum black holes,''
  arXiv:0806.2051 [hep-th].

\bibitem{KLR2}
  H.~K.~Kunduri, J.~Lucietti and H.~S.~Reall,
  ``Near-horizon symmetries of extremal black holes,''
  Class.\ Quant.\ Grav.\  {\bf 24} (2007) 4169
  [arXiv:0705.4214 [hep-th]].

\bibitem{KL4d}
H.~K.~Kunduri and J.~Lucietti,
  ``Uniqueness of near-horizon geometries of rotating extremal AdS(4) black
  holes,''
  Class.\ Quant.\ Grav.\  {\bf 26}, 055019 (2009)
  [arXiv:0812.1576 [hep-th]].

\bibitem{LP}
  J.~Lewandowski and T.~Pawlowski,
  ``Extremal Isolated Horizons: A Local Uniqueness Theorem,''
  Class.\ Quant.\ Grav.\  {\bf 20} (2003) 587
  [arXiv:gr-qc/0208032].

\bibitem{CRT2}
  P.~T.~Chrusciel, H.~S.~Reall and P.~Tod,
  ``On Israel-Wilson-Perjes black holes,''
  Class.\ Quant.\ Grav.\  {\bf 23} (2006) 2519
  [arXiv:gr-qc/0512116].

\bibitem{KPTod}
  K.~p.~Tod,
  ``All Metrics Admitting Supercovariantly Constant Spinors,''
  Phys.\ Lett.\  B {\bf 121}, 241 (1983).

\bibitem{CRT1}
  P.~T.~Chrusciel, H.~S.~Reall and P.~Tod,
  ``On non-existence of static vacuum black holes with degenerate  components
  of the event horizon,''
  Class.\ Quant.\ Grav.\  {\bf 23} (2006) 549
  [arXiv:gr-qc/0512041].

\bibitem{Lap}
S.~Rosenberg,
  ``The Laplacian on a Riemannian Manifold,''
{\it  London Mathematical Society, Cambridge University Press} (1997) 174p.


\bibitem{KLRnoring}
  H.~K.~Kunduri, J.~Lucietti and H.~S.~Reall,
  ``Do supersymmetric anti-de Sitter black rings exist?,''
  JHEP {\bf 0702} (2007) 026
  [arXiv:hep-th/0611351].

\bibitem{Chr}
  P.~Chrusciel,
  ``On space-times with $U(1)\times U(1)$ symmetric compact Cauchy surfaces,''
  Annals Phys.\  {\bf 202} (1990) 100.

\bibitem{Gowdy}
  R.~H.~Gowdy,
  ``Vacuum space-times with two parameter spacelike isometry groups and compact invariant hypersurfaces: Topologies and boundary conditions,''
  Annals Phys.\  {\bf 83} (1974) 203.

\bibitem{EMDring}
  H.~K.~Kunduri and J.~Lucietti,
  ``Electrically charged dilatonic black rings,''
  Phys.\ Lett.\  B {\bf 609} (2005) 143
  [arXiv:hep-th/0412153].

\bibitem{Gal}
  D.~V.~Gal'tsov and O.~A.~Rytchkov,
  ``Generating branes via sigma-models,''
  Phys.\ Rev.\  D {\bf 58} (1998) 122001
  [arXiv:hep-th/9801160].

\bibitem{Bena}
  I.~Bena,
  ``Splitting hairs of the three charge black hole,''
  Phys.\ Rev.\  D {\bf 70} (2004) 105018
  [arXiv:hep-th/0404073].

\bibitem{Emp}
  R.~Emparan,
  ``Tubular branes in fluxbranes,''
  Nucl.\ Phys.\  B {\bf 610} (2001) 169
  [arXiv:hep-th/0105062].

}
\end{thebibliography}
\end{document}